\documentclass[floatfix,
reprint,
superscriptaddress,
amsmath,amssymb,
aps,
prl,
]{revtex4-2}

\usepackage{graphicx}
\usepackage[dvipsnames]{xcolor}
\usepackage{pdfpages}
\usepackage{amssymb}
\usepackage{amsmath}
\usepackage{amsbsy}
\usepackage{color}
\usepackage[normalem]{ulem}
\usepackage{epstopdf}
\usepackage{bm}
\usepackage{xurl}
\usepackage{textcomp}

\usepackage{float}

\usepackage{filecontents}
\makeatletter\@input{z2.tex}\makeatother

\newcommand{\ldna}{$\lambda$DNA~}

\graphicspath{{figures/}}

\usepackage{xr}
\makeatletter
\AtBeginDocument{\let\LS@rot\@undefined}
\newcommand*{\addFileDependency}[1]{
  \typeout{(#1)}
  \@addtofilelist{#1}
  \IfFileExists{#1}{}{\typeout{No file #1.}}
}
\makeatother
\newcommand*{\myexternaldocument}[1]{%
    \externaldocument[SM-]{#1}%
    \addFileDependency{#1.tex}%
    \addFileDependency{#1.aux}%
}
\myexternaldocument{si}

\begin{document}
\title{Modulation of DNA rheology by a transcription factor that forms aging microgels}

\author{Amandine Hong-Minh}
\thanks{Joint first author}
\affiliation{Centre for Regenerative Medicine, Institute for Regeneration and Repair, 5 Little France Drive, Edinburgh EH16 4UU, Scotland}
\affiliation{School of Physics and Astronomy, University of Edinburgh, Edinburgh Eh9 3FD, Scotland}

\author{Yair Augusto Guti\'{e}rrez Fosado}
\thanks{Joint first author}
\affiliation{School of Physics and Astronomy, University of Edinburgh, Edinburgh Eh9 3FD, Scotland}

\author{Abbie Guild}
\affiliation{Centre for Regenerative Medicine, Institute for Regeneration and Repair, 5 Little France Drive, Edinburgh EH16 4UU, Scotland}
\affiliation{Institute for Stem Cell Research, School of Biological Sciences, University of Edinburgh, 5 Little France Drive, Edinburgh EH16 4UU, Scotland}

\author{Nicholas Mullin}
\affiliation{Centre for Regenerative Medicine, Institute for Regeneration and Repair, 5 Little France Drive, Edinburgh EH16 4UU, Scotland}
\affiliation{Institute for Stem Cell Research, School of Biological Sciences, University of Edinburgh, 5 Little France Drive, Edinburgh EH16 4UU, Scotland}

\author{Laura Spagnolo}
\affiliation{School of Molecular Biosciences, University of Glasgow, University Avenue, Glasgow G12 8QQ, Scotland}

\author{Ian Chambers}
\thanks{corresponding author, ichambers@ed.ac.uk}
\affiliation{Centre for Regenerative Medicine, Institute for Regeneration and Repair, 5 Little France Drive, Edinburgh EH16 4UU, Scotland}
\affiliation{Institute for Stem Cell Research, School of Biological Sciences, University of Edinburgh, 5 Little France Drive, Edinburgh EH16 4UU, Scotland}

\author{Davide Michieletto}
\thanks{corresponding author, davide.michieletto@ed.ac.uk}
\affiliation{School of Physics and Astronomy, University of Edinburgh, Edinburgh Eh9 3FD, Scotland}
\affiliation{MRC Human Genetics Unit, Institute of Genetics and Cancer, University of Edinburgh, Scotland}
\affiliation{International Institute for Sustainability with Knotted Chiral Meta Matter (WPI-SKCM$^2$), Hiroshima University, Higashi-Hiroshima, Hiroshima 739-8526, Japan}

\begin{abstract}
\textbf{Proteins and nucleic acids form non-Newtonian liquids with complex rheological properties that contribute to their function \textit{in vivo}. Here we investigate the rheology of the transcription factor NANOG, a key protein in sustaining embryonic stem cell self-renewal. We discover that at high concentrations NANOG forms macroscopic aging gels through its intrinsically disordered tryptophan-rich domain. By combining molecular dynamics simulations, mass photometry and Cryo-EM, we also discover that NANOG forms self-limiting micelle-like clusters which expose their DNA-binding domains. In dense solutions of DNA, NANOG micelle-like structures stabilize inter-molecular entanglements and crosslinks, forming microgel-like structures. Our findings suggest that NANOG may contribute to regulate gene expression in a unconventional way: by restricting and stabilizing genome dynamics at key transcriptional sites through the formation of an aging microgel-like structure, potentially enabling mechanical memory in the gene network.  
}
\end{abstract}

\maketitle

\begin{figure*}[t]
	\begin{center}		\includegraphics[width=0.95\textwidth]{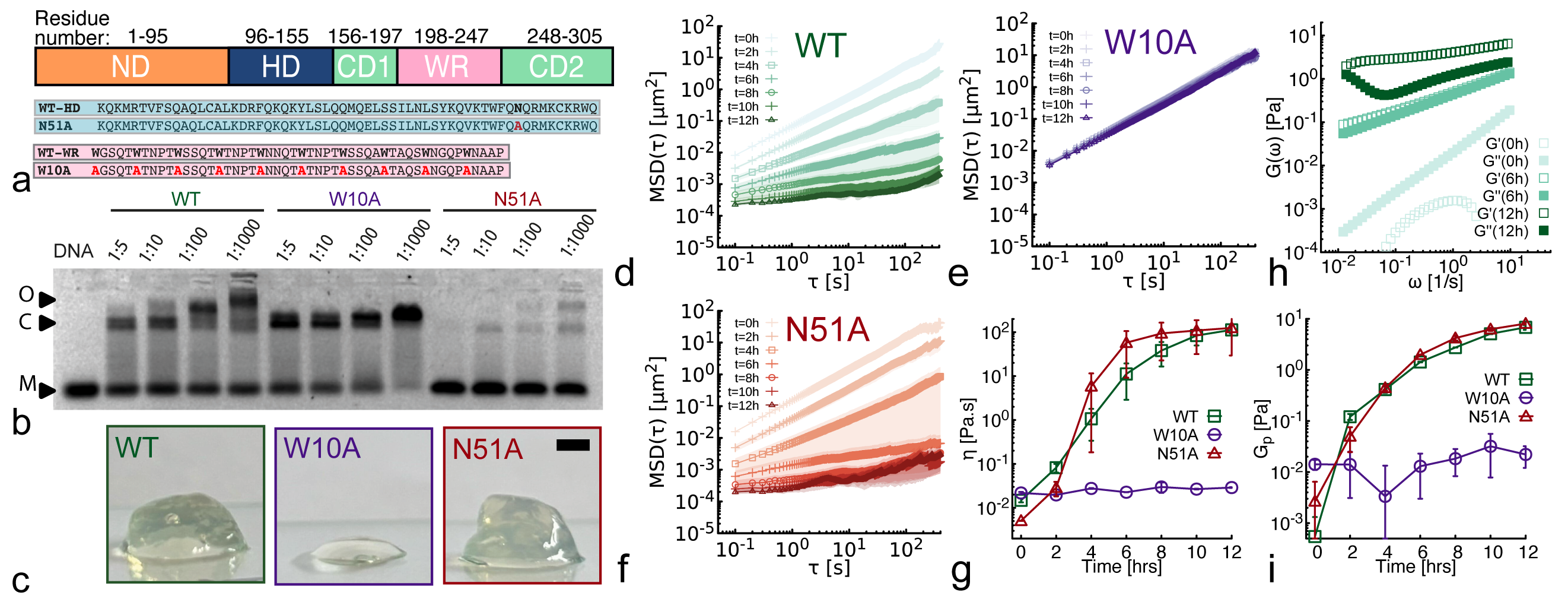}
        \caption{\textbf{NANOG solutions form macroscopic aging gels through the intrinsically disordered WR domain.} \textbf{a.} Mouse NANOG primary structure: N-terminal domain (ND), homeodomain (HD), C-terminal domains (CD1,2), tryptophan repeat (WR). Amino acid sequences of HD and WR mutants are shown. WT residues are shown in black; mutated residues are shown in red. \textbf{b.} EMSA of NANOG mutants at different DNA:protein ratios. Each lane contains 5 nM of a 26-bp dsDNA oligomer. (M=monomer, C=DNA-NANOG complex, O=oligomer). \textbf{c.} Photographs of $\sim$mL samples of purified NANOG mutants aged at  $\sim$1 mM overnight at 37$^\circ$C. WT NANOG and N51A form a non-pipettable gel, whereas W10A remains liquid. Scale bar is 2 mm. \textbf{d-e-f.} Microrheology of the 3 mutants over the course of 12h at 37$^\circ$C. \textbf{g.} The samples' viscosity showing a 10'000-fold increase for WT and N51A and absence of aging for the W10A mutant lacking the tryptophan residues. \textbf{h.} Elastic ($G^\prime$) and viscous ($G^{\prime \prime}$) moduli for 3 ageing times. \textbf{i.} Elasticity of the sample taken as $G_p = G^\prime(\omega=10$ Hz) showing a significant gelation for the WT and N51A proteins over the course of 12 hours, while W10A displays no increase in elasticity. }
        \label{fig:aging_protein}
	\end{center}
    \vspace{-0.6cm}
\end{figure*}

Cell fate is regulated by transcription factors (TFs), specialized proteins that bind the genome at key regulatory sites and regulate gene expression~\cite{alberts2022molecular}. In embryonic stem cells (ESC), three TFs -- OCT4, SOX2, and NANOG -- represent the core regulatory network needed to maintain a pluripotent state and prevent differentiation~\cite{Niwa2000,Chambers2003,Masui2007,Mitsui2003Nanog,KarwackiNeisius2013}. 
Among these, NANOG is unique in its ability to sustain ESC self-renewal in the absence of the otherwise essential Leukemia Inhibitory Factor~\cite{Chambers2007}. However, the precise physical mechanisms enabling NANOG to regulate cell fate are not known. 

The intrinsically disordered region of the protein (Fig.~\ref{fig:aging_protein}a) -- referred to as tryptophan repeat (WR) -- enables it to form oligomers~\cite{Mullin2008,Wang2008} and recent work \textit{in vitro} and in non-pluripotent cells has further suggested that DNA-bridging~\cite{Choi2022} and liquid-liquid phase separation~\cite{Boija2018,McSwiggen2019} may contribute to the cellular function of NANOG. When expressed in ESC, NANOG variants that display mutations in either the WR domain (W10A mutant, Fig.~\ref{fig:aging_protein}b) or the DNA binding homeodomain (N51A, fig.~\ref{fig:aging_protein}b), both fail to maintain pluripotency (see Refs.~\cite{Mullin2017,Novo2016Nanog} and unpublished data). It is currently conjectured that the combination of WR and DNA-binding domain enable NANOG to bring together distant DNA elements, e.g. enhancer and promoters~\cite{Novo2018LongRange}. However there is no evidence of major genome spatial re-organization following NANOG over-expression \textit{in vivo}, suggesting that NANOG oligomerization does not lead to significant spatial genome remodeling~\cite{deWit2013Pluripotent}. It thus remains poorly understood why the WR domain is essential for ESC self-renewal. 

Although NANOG was found to form coacervates with DNA and other proteins, such as Mediator complex and SOX2 both \textit{in vivo} and \textit{in vitro}~\cite{Boija2018,Gagliardi2013}, these interactions are likely enabled by its WR. While this evidence underscore the potential role of the NANOG WR to form multivalent interactions~\cite{Mullin2017} and recruit gene expression activators at key regulatory sites in the genome, they do not explain why NANOG oligomerization is necessary for ESC renewal.

Given the lack of large scale reorganization of the genome upon overexpression of NANOG~\cite{Choi2022,deWit2013Pluripotent}, we hypothesized that the NANOG oligomerization domain (WR) may contribute to regulate DNA dynamics. We thus decided to characterize the rheology of NANOG and DNA-NANOG complex fluids \textit{in vitro}. To do this, we have combined microrheology, cryo-electron microscopy, and computational simulations to characterize the material properties and molecular organization of NANOG complex fluids. We discovered that NANOG forms self-limited micelle-like structures with surface-exposed DNA-binding domains, thus acting as a transient cross-linker for DNA molecules. We further observe a significant aging of these complex fluids, which align with the behavior of other intrinsically disordered RNA-binding proteins~\cite{Jawerth2020}. Our observations suggest that NANOG may contribute to modulate gene expression at key regulatory sites in a unconventional way, i.e. by creating transient multivalent cross-links which become progressively strong during aging of the micelle-like structure, thus providing a physical way to ingrain memory in a regulatory gene network. 

\begin{figure*}[t!]
	\begin{center}
		\includegraphics[width=1.0\textwidth]{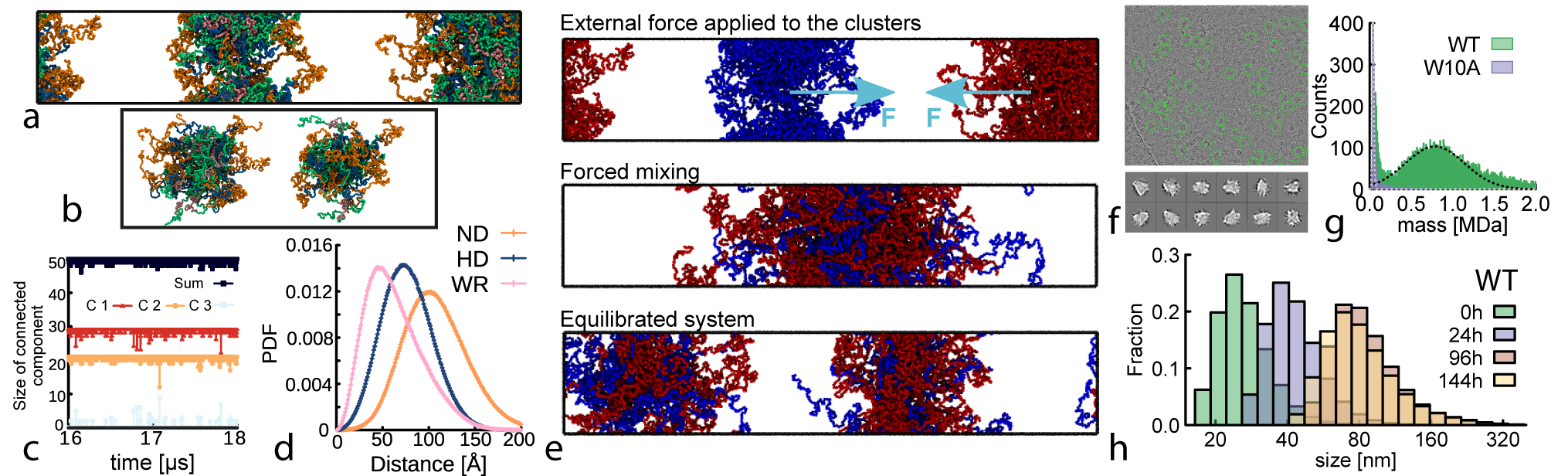}
		\caption{\textbf{NANOG forms micelle-like self-limited clusters through their WR}. \textbf{a.} Snapshot from equilibrated simulations of NANOG made of $M = 50$ proteins, each $N = 305$ amino acids long at $T = 300~\mathrm{K}$ and $\rho = 0.1~\mathrm{g/cm^3}$. Green, orange, pink and blue represent the different NANOG domains as in Fig.~\ref{fig:aging_protein}a. \textbf{b.} Reconstruction of the system in (a) through periodic boundary conditions showing two disjoint clusters. \textbf{c.} Time evolution of the number of molecules forming the two main clusters and the third smaller cluster of molecules. \textbf{d.} Distribution of distances between the cluster COM to the residues in the different domains: N-terminal domain (ND, orange), homeodomain (HD, blue) and  tryptophan repeat (WR, pink). \textbf{e.} Simulated forced mixing of clusters (colored red and blue) and re-equilibration leading to the re-establishment of two disjoint clusters made of a different combination of molecules.  
        \textbf{f.} Representative cryo-EM motion corrected micrographs (scale bar 50 nm). \textbf{g.} 2D classes of objects showing highly heterogeneous and ``disordered'' assembly. Box size is 364 \AA, mask is 240 \AA, approximately the size of $\sim$20-30 NANOG molecules. \textbf{g.} Mass photometry showing peaks at weight corresponding to $\sim$22-25 NANOGs per cluster. \textbf{h.} DLS quantifying the hydrodynamic size of clusters in samples of NANOG incubated at 37$^\circ$C and different times.
        }
        \label{fig2}
	\end{center}
    \vspace{-0.6cm}
\end{figure*}

\paragraph{Results --}
Inspired by recent observations that intrinsically disordered proteins display glass-like aging~\cite{Jawerth2020}, we decided to investigate the biophysical behavior of NANOG -- a TF possessing an intrinsically disordered $\sim$50 amino acid domain -- and that of two of its mutants (see Fig.~\ref{fig:aging_protein}a): (i) a N51A mutant carrying a substitution in the homeodomain that significantly decreases DNA binding and (ii) a W10A mutant in which all ten tryptophans in the WR are replaced by alanines. Previous work demonstrated that the W10A mutant fails to oligomerize~\cite{Mullin2017,Wang2008} and that both N51A and W10A lack the ability to maintain ESC self-renewal when overexpressed ~\cite{Navarro2012NanogAutorepression,Mullin2017}. 

To quantify DNA binding and oligomerization, we first performed an electrophoretic mobility shift assay (EMSA). Here, recombinant NANOG protein was incubated with a 26 bp DNA fragment from the mouse tcf promoter known to be specifically bound by WT NANOG~\cite{Jauch2008} and then run on a gel. Since nucleoprotein complexes migrate slower than free DNA through the gel, we can visualize DNA binding and oligomerization ability of various NANOG mutants. In Figure~\ref{fig:aging_protein}b we show that the N51A mutant exhibits substantially reduced DNA-binding affinity, while W10A lacks the oligomeric band formed at high protein concentration, as expected. We further tested whether NANOG displayed aging material properties  similar to those of other intrinsically disordered proteins~\cite{Jawerth2020,Michieletto2022}. We thus incubated 1mM NANOG WT, N51A and W10A at 37$^\circ$C overnight and noticed that both WT and N51A underwent a sol-gel transition while W10A remained liquid (Figure~\ref{fig:aging_protein}c). Performing the same incubation at room temperature did not trigger the phase transition in any of the proteins. 

We more precisely quantified the timescales of gel formation by performing microrheology, i.e. by measuring the mean squared displacement (MSD) of passive tracers embedded in the aging solutions as $\langle \Delta r^2 (\tau) \rangle = \langle \bm{r}(t+\tau) - \bm{r}(t) \rangle$. As shown in Fig.~\ref{fig:aging_protein}d,f tracers in both WT and N51A started to display subdiffusive behavior ($\langle \Delta r^2 \rangle \sim t^\alpha$ with $\alpha<1$) within 4-6 hours of incubation at 37$^\circ$C. The tracers' subdiffusive behavior indicates the onset of elasticity and gelling~\cite{Michieletto2022,Cicuta2007}. On the other hand, tracers embedded within the W10A solution remained freely diffusive ($\langle \Delta r^2 \rangle \sim t^1$, Fig.~\ref{fig:aging_protein}e). This suggests that loss of tryptophans in the WR, which abolish self-interactions, also prevented gel formation. The formation of a macroscopic gel suggests that the WR mediates multivalent strong interactions between NANOG proteins. After 12h of incubation the effective viscosity of the WT and N51A systems, obtained as $\eta = \lim_{t \to \infty} \langle \Delta r^2 (t)\rangle/6t$ was $10^4$-fold larger than the solution pre-incubation (Fig.~\ref{fig:aging_protein}g). We further characterized the aging through the generalized Stokes-Einstein relation (GSER)~\cite{Mason2000,Mason1995,Cicuta2007,Brizioli2025OFM}, $G^*(\omega) = (k_B T)/(\pi a i \omega \mathcal{F}_u \left[ \langle \Delta r^2 \rangle \right])$, which allows us to compute the elastic and viscous shear moduli $G^\prime(\omega)$ and $G^{\prime \prime}(\omega)$. In Fig.~\ref{fig:aging_protein}h, we show that at around 6h, the elastic modulus start to dominate over the viscous one, in turn identifying this as the point of gelation. The elasticity of the WT and N51A samples also grow $10^4$-fold over 12 hours, as shown in Fig.~\ref{fig:aging_protein}i, however W10A displays no elastic response. Importantly, the fact that we do not observe sol-gel transition at room temperature in any protein sample indicates that the aging process is driven by WR re-configuration and exploration of a rugged free energy landscape within NANOG oligomers, which is facilitated by higher temperatures. While this argument is in line with what expected for aging Maxwell fluids~\cite{Jawerth2020}, our time-cure superposition analysis does not agree with the one expected for simple Maxwell fluids and instead display multiple characteristic timescales (see SI for more details).

To better understand the gel formation and its biological relevance, we decided to simulate NANOG proteins using the Mpipi framework: a residue-level model parameterized on both bioinformatics data and all-atom simulations~\cite{Mpipi2021}. We simulated $M = 50$ NANOG proteins, each consisting of $N = 305$ amino acids. The proteins were initially uniformly distributed in an elongated simulation box corresponding to an overall density of $\rho = 0.1~\mathrm{g/cm^3}$. The system was then evolved in the NVT ensemble using LAMMPS~\cite{LAMMPS2022} with implicit solvent at a temperature of $T = 300$ K. A time step of $10~\mathrm{fs}$ was used for integration over approximately $2 \times 10^9$ steps (i.e., $20~\mu$s) and we ran 3 independent replicas (see SI for further details). Our simulations revealed that NANOG proteins assemble into self-limited clusters, each characterized by a well-defined maximum size of approximately 30 proteins. Figures~\ref{fig2}a-b show representative snapshots depicting two stable clusters. Although the clusters' composition was dynamic (Fig.~\ref{fig2}c), the overall number of molecules per cluster remained the same through the simulation. These observations suggest that there is exchange of proteins between clusters, however they are self-limited. To further characterize the internal structure of clusters we computed the center of mass (COM) of the clusters and measured the distances of all residues in the different regions (ND, HD, WR) from the cluster COM. The resulting distributions (Fig.~\ref{fig2}d) show that NANOG molecules form a micelle-like structure where the WR occupies the cluster center, the HD (DNA binding domain) occupies an intermediate layer, while the ND forms the outer layer of the micelle-like cluster. Similar micelle-like organization was observed in simulations of human NANOG, however the observed cluster size was larger~\cite{Takada2023}.

\begin{figure}[t!]
	\begin{center}
\includegraphics[width=0.49\textwidth]{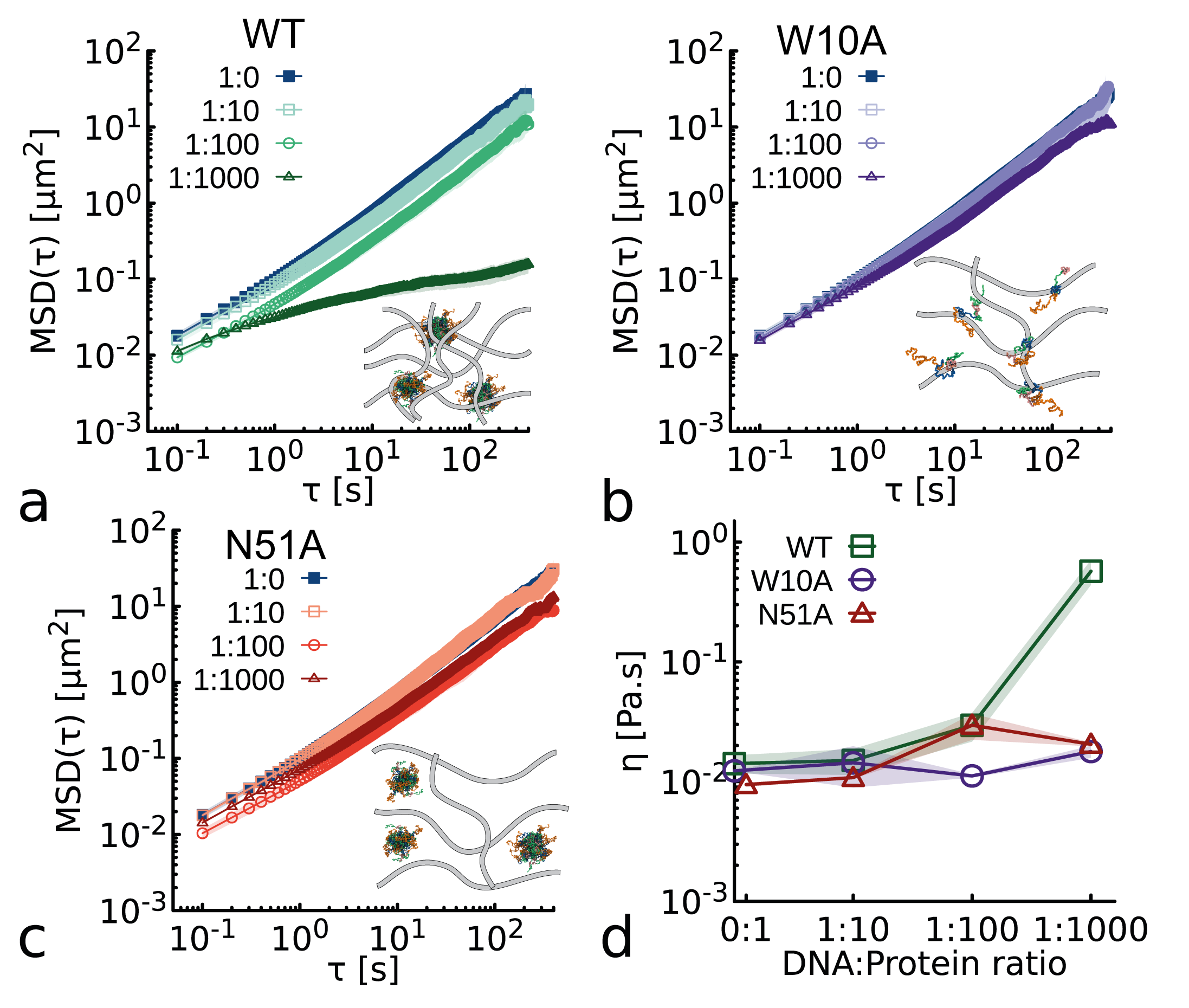}
		\caption{\textbf{NANOG micelle-like clusters act as crosslinkers of entangled DNA}. 
        \textbf{a-c.} MSD of tracers at varying DNA:protein stoichiometries in a solution of \ldna at 7.9~nM and (a) WT, (b) W10A and (c) N51A, at different concentrations. Insets depict schematics of NANOG structures in the presence of \ldna. \textbf{d.} Viscosity of DNA–protein solutions as a function of protein concentration. }
        \label{fig3}
	\end{center}
    \vspace{-1cm}
\end{figure}

To further test the stability of the clusters we artificially forced mixing of the two clusters (marked red and blue in Fig.~\ref{fig2}e, top). Once the clusters merged, we removed the external mixing forces and the system was allowed to re-equilibrate for $20\,\mu\text{s}$. Strikingly, we observed the re-organization of the proteins into two distinct clusters formed by two new subsets of proteins (see Fig.~\ref{fig2}e). This behavior demonstrates that NANOG sequence possess properties akin to block co-polymers, which can thermodynamically display stable phases with finite-size micelle-like assemblies~\cite{Hamley2007}.

To experimentally test our simulations we performed cryo-EM and mass photometry on systems consisting of purified NANOG WT and W10A. Cryo-EM identified a range of heterogeneous and amorphous structures (Fig.~\ref{fig2}f-g), displaying an overall similar shape and size (approximately 25 nm in diameter) across the dataset. Yet structures did not lock into a discrete conformational space that would allow high resolution reconstruction. In fact, the objects are rather blurry, suggesting significant local resolution variation typical of intrinsically disordered proteins. Further, mass photometry identified clusters made of $\sim$ 22-25 NANOG monomers. Performing the same experiments on W10A could not identify any oligomers; accordingly, our simulations of W10A mutant found no onset of clustering (see SI). Finally, to experimentally test self-limiting growth, we performed Dynamic Light Scattering (Zetasizer Pro, Malvern Instruments) measurements at different aging times, during 37$^\circ$C incubation of 1 mg/ml NANOG sample. The distributions in Fig.~\ref{fig2}h display a clear shift towards larger hydrodynamic size with time, however the shift stops after 96h incubation, confirming the self-limiting nature of the NANOG clusters.

To better understand the physiological role of NANOG gelling, we performed microrheology~\cite{Mason1995,Zhu2008} on solutions made by 250 ng/$\mu$l of \ldna and varying protein concentrations. Note that $c = 250$ ng/$\mu$l is around 10-fold larger than \ldna overlapping concentration $c^* \simeq 20$ ng/$\mu$l and the DNA is therefore entangled.
As shown in Figs.~\ref{fig3}a-c, WT NANOG displays a marked reduction in tracer mobility at around 7.9 $\mu$M concentration. In contrast, W10A and N51A display no significant change in MSDs over the explored concentration range.
We rationalize this difference as follows: WT NANOG forms micelle-like clusters where the DNA-binding domains is exposed at the micelle surface, enabling each cluster to bridge multiple \ldna molecules and in turn modulating the solution's rheology (Fig.~\ref{fig3}d). W10A does not form clusters but still binds DNA through the homeodomain, whereas N51A forms clusters but has a much weaker DNA binding. In both these cases, the lack of simultaneous clustering and DNA binding prevents DNA bridging, and consequently, no major effect on rheology is observed.
This behaviour is akin to that of other DNA-protein condensates which show solid-like behaviour with long DNA~\cite{Muzzopappa2021,Michieletto2022}.

\begin{figure}[t!]
	\begin{center} \includegraphics[width=0.45\textwidth]{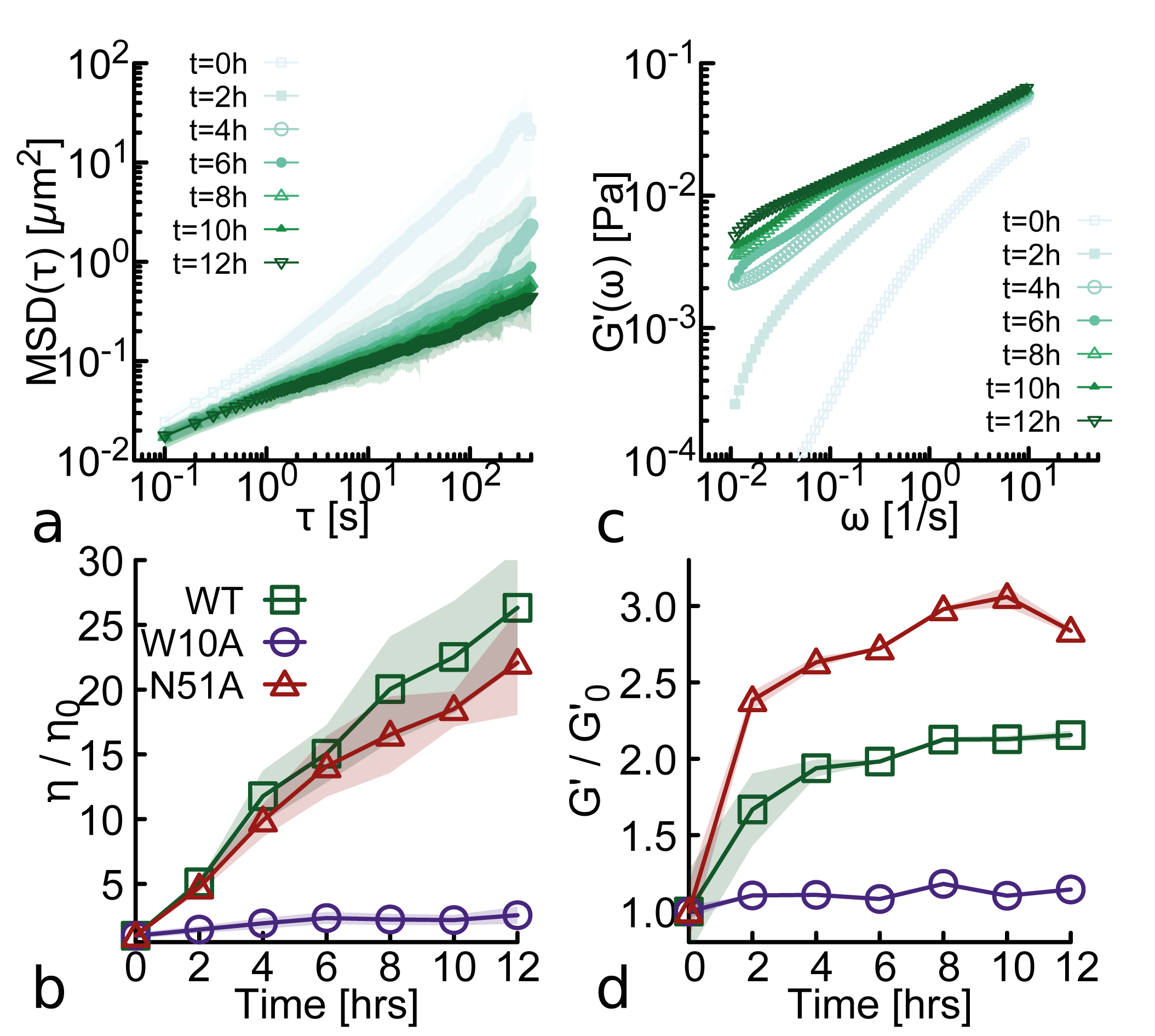}
		\caption{\textbf{Aging of NANOG-DNA solutions suggest a  rheological gene regulation}. 
        \textbf{a.} Time resolved MSDs of tracers' embedded in solutions of NANOG (7.9~$\mu$M) and \ldna (7.9 nM) for 12 hours at 37$^{\circ}$C. \textbf{b.} Normalized viscosity of \ldna-NANOG solutions obtained for different mutants. \textbf{c.} Time resolved elastic modulus $G^\prime$ obtained from the MSD via the GSER. \textbf{d.} Solution's elasticity $G_p = G^\prime (\omega=10 \textrm{Hz})$ as a function of ageing time. } 
        \label{fig4}
	\end{center}
    \vspace{-0.6cm}
\end{figure}

Finally, we investigated mixtures of entangled \ldna (7.9nM) with each of the NANOG variants (7.9 $\mu$M), corresponding to a 1:1000 DNA-to-protein molar ratio. Despite the lower protein concentration compared with Fig.~\ref{fig:aging_protein} (where 1 mM was used), we still observe aging of the samples. In fact, both WT and N51A samples mixed with \ldna display aging and increase in viscosity (Fig.~\ref{fig4}a-b) and elasticity (Fig.~\ref{fig4}c-d). 
The fact that both WT and N51A age suggest that even in the presence of weak DNA binding (N51A), strong oligomerization is sufficient to drive slow structural rearrangements toward a viscoelastic state. By contrast, W10A lacks this oligomerization capability and therefore remains in a viscous-dominated state.

\paragraph*{Conclusions--} We combined simulations, cryo-EM, mass photometry, DLS and microrheology to dissect the molecular and rheological properties of a powerful transcription factor, NANOG, and of NANOG-DNA fluids. By comparing wild-type NANOG with two mutants -- W10A, which cannot oligomerize, and N51A, which has weak DNA binding -- we uncovered the distinct contributions of the intrinsically disordered domain (WR) and the DNA binding domain (HD) to NANOG’s self-assembly and DNA interactions. 

First, we discovered that NANOG WT is able to form an aging macroscopic gel (Fig.~\ref{fig:aging_protein}); we then performed residue-level simulations which revealed that NANOG forms self-limited clusters, a prediction confirmed by cryo-EM, mass photometry and DLS (Fig.~\ref{fig2}). Simulations suggest that these micelle-like clusters expose the DNA-binding domain, providing a physical mechanism for DNA bridging. At high concentration, these micelle-like structures cross-link and drive the formation of a microgel, i.e. a gel formed by small ($\sim$ nm-$\mu$m) sticky domains.

Finally, microrheology experiments further demonstrated (Fig.~\ref{fig3}) that WT NANOG–DNA mixtures exhibit markedly enhanced viscoelasticity compared to either mutants, consistent with a requirement for both clustering and DNA binding to achieve DNA bridging. Importantly, solutions of NANOG variants also display aging (Fig.~\ref{fig4}), whereby viscoelasticity increases over time. Aging occurs in WT and N51A, but not in W10A, indicating that oligomerization is sufficient to drive slow restructuring toward gel-like states, while DNA bridging further accelerates this process.  

We speculate that the gel-like material properties of NANOG solutions may provide a physical basis for stabilizing long-range chromatin interactions and enhancer–promoter contacts, thereby contributing to transcriptional regulation in pluripotent cells. However, there is no evidence suggesting that NANOG overexpression drives long-range chromatin reorganization. We thus conjecture that NANOG may contribute to regulate gene expression in an unconventional way, i.e. by stabilizing and restricting the \textit{dynamics} of key regulatory sites in such a way that they are stably up/down regulated. This so far unexplored mechanism could also explain liquid-to-solid phase transitions observed in other TFs~\cite{Pulupa2025}. The formation of aging condensates assuming solid-like properties over time may reinforce, stabilize and ingrain memory in a regulatory network, including enhancer-promoter interactions.   

Whether NANOG micelle-like structures and gel-like properties influence biological function remains an open question. For instance, it would be interesting to simultaneously track fluorescently labeled NANOG and chromatin loci upon overexpression of NANOG in living mouse ESC. Our hypothesis is that areas where NANOG is enriched should correlate with slow chromatin dynamics. This hypothesis, if true, would reveal a new paradigm through which biomolecular condensates regulate genome organization and gene expression, i.e. by regulating the dynamics of key regulatory genomic sites rather than genome folding.  

More generally, in the future it will also be interesting to investigate the combined effect of NANOG with different TFs, such as Sox2, Oct4, and Klf4~\cite{Takada2025}, to study other gel-forming and RNA-binding nuclear proteins~\cite{Michieletto2022,Marenda2024} and their influence on chromatin structure and dynamics~\cite{Li2025Chromatin}.

\section{Acknowledgements}
YAGF acknowledges support from the Physics of Life, UKRI/Wellcome (grant number EP/T022000/1–PoLNET3). DM acknowledges the Royal Society and the European Research Council (grant agreement No 947918, TAP) for funding. IC acknowledges grant support from the MRC (MR/T003162/1) and BBSRC (BB/T008644/1). We also acknowledge support from MRC Precision Medicine PhD programme and Wellcome Trust integrated cellular mechanisms PhD programme. 
We acknowledge the Scottish Centre for Macromolecular Imaging, Mhairi-Clarke and James-Streetley for assistance with cryo-EM experiments and access to instrumentation, financed by the Medical Research Council (MC-PC-17135) and the Scottish Funding Council (H17007). For the purpose of open access, we have applied a Creative Commons Attribution (CCBY) licence to any author accepted manuscript version arising from this submission.

\bibliography{bibliography}

\begin{thebibliography}{35}%
\makeatletter
\providecommand \@ifxundefined [1]{%
 \@ifx{#1\undefined}
}%
\providecommand \@ifnum [1]{%
 \ifnum #1\expandafter \@firstoftwo
 \else \expandafter \@secondoftwo
 \fi
}%
\providecommand \@ifx [1]{%
 \ifx #1\expandafter \@firstoftwo
 \else \expandafter \@secondoftwo
 \fi
}%
\providecommand \natexlab [1]{#1}%
\providecommand \enquote  [1]{``#1''}%
\providecommand \bibnamefont  [1]{#1}%
\providecommand \bibfnamefont [1]{#1}%
\providecommand \citenamefont [1]{#1}%
\providecommand \href@noop [0]{\@secondoftwo}%
\providecommand \href [0]{\begingroup \@sanitize@url \@href}%
\providecommand \@href[1]{\@@startlink{#1}\@@href}%
\providecommand \@@href[1]{\endgroup#1\@@endlink}%
\providecommand \@sanitize@url [0]{\catcode `\\12\catcode `\$12\catcode
  `\&12\catcode `\#12\catcode `\^12\catcode `\_12\catcode `\%12\relax}%
\providecommand \@@startlink[1]{}%
\providecommand \@@endlink[0]{}%
\providecommand \url  [0]{\begingroup\@sanitize@url \@url }%
\providecommand \@url [1]{\endgroup\@href {#1}{\urlprefix }}%
\providecommand \urlprefix  [0]{URL }%
\providecommand \Eprint [0]{\href }%
\providecommand \doibase [0]{https://doi.org/}%
\providecommand \selectlanguage [0]{\@gobble}%
\providecommand \bibinfo  [0]{\@secondoftwo}%
\providecommand \bibfield  [0]{\@secondoftwo}%
\providecommand \translation [1]{[#1]}%
\providecommand \BibitemOpen [0]{}%
\providecommand \bibitemStop [0]{}%
\providecommand \bibitemNoStop [0]{.\EOS\space}%
\providecommand \EOS [0]{\spacefactor3000\relax}%
\providecommand \BibitemShut  [1]{\csname bibitem#1\endcsname}%
\let\auto@bib@innerbib\@empty
\bibitem [{\citenamefont {Alberts}\ \emph {et~al.}(2022)\citenamefont
  {Alberts}, \citenamefont {Heald}, \citenamefont {Johnson}, \citenamefont
  {Morgan}, \citenamefont {Raff}, \citenamefont {Roberts},\ and\ \citenamefont
  {Walter}}]{alberts2022molecular}%
  \BibitemOpen
  \bibfield  {author} {\bibinfo {author} {\bibfnamefont {B.}~\bibnamefont
  {Alberts}}, \bibinfo {author} {\bibfnamefont {R.}~\bibnamefont {Heald}},
  \bibinfo {author} {\bibfnamefont {A.}~\bibnamefont {Johnson}}, \bibinfo
  {author} {\bibfnamefont {D.}~\bibnamefont {Morgan}}, \bibinfo {author}
  {\bibfnamefont {M.}~\bibnamefont {Raff}}, \bibinfo {author} {\bibfnamefont
  {K.}~\bibnamefont {Roberts}},\ and\ \bibinfo {author} {\bibfnamefont
  {P.}~\bibnamefont {Walter}},\ }\href@noop {} {\emph {\bibinfo {title}
  {Molecular Biology of the Cell}}},\ \bibinfo {edition} {7th}\ ed.\ (\bibinfo
  {publisher} {W. W. Norton \& Company},\ \bibinfo {year} {2022})\BibitemShut
  {NoStop}%
\bibitem [{\citenamefont {Niwa}\ \emph {et~al.}(2000)\citenamefont {Niwa},
  \citenamefont {Miyazaki},\ and\ \citenamefont {Smith}}]{Niwa2000}%
  \BibitemOpen
  \bibfield  {author} {\bibinfo {author} {\bibfnamefont {H.}~\bibnamefont
  {Niwa}}, \bibinfo {author} {\bibfnamefont {J.-i.}\ \bibnamefont {Miyazaki}},\
  and\ \bibinfo {author} {\bibfnamefont {A.~G.}\ \bibnamefont {Smith}},\ }\href
  {https://doi.org/10.1038/74199} {\bibfield  {journal} {\bibinfo  {journal}
  {Nature Genetics}\ }\textbf {\bibinfo {volume} {24}},\ \bibinfo {pages} {372}
  (\bibinfo {year} {2000})}\BibitemShut {NoStop}%
\bibitem [{\citenamefont {Chambers}\ \emph {et~al.}(2003)\citenamefont
  {Chambers}, \citenamefont {Colby}, \citenamefont {Robertson}, \citenamefont
  {Nichols}, \citenamefont {Lee}, \citenamefont {Tweedie},\ and\ \citenamefont
  {Smith}}]{Chambers2003}%
  \BibitemOpen
  \bibfield  {author} {\bibinfo {author} {\bibfnamefont {I.}~\bibnamefont
  {Chambers}}, \bibinfo {author} {\bibfnamefont {D.}~\bibnamefont {Colby}},
  \bibinfo {author} {\bibfnamefont {M.}~\bibnamefont {Robertson}}, \bibinfo
  {author} {\bibfnamefont {J.}~\bibnamefont {Nichols}}, \bibinfo {author}
  {\bibfnamefont {S.}~\bibnamefont {Lee}}, \bibinfo {author} {\bibfnamefont
  {S.}~\bibnamefont {Tweedie}},\ and\ \bibinfo {author} {\bibfnamefont
  {A.}~\bibnamefont {Smith}},\ }\href@noop {} {\bibfield  {journal} {\bibinfo
  {journal} {Cell}\ }\textbf {\bibinfo {volume} {113}},\ \bibinfo {pages} {643}
  (\bibinfo {year} {2003})}\BibitemShut {NoStop}%
\bibitem [{\citenamefont {Masui}\ \emph {et~al.}(2007)\citenamefont {Masui},
  \citenamefont {Nakatake}, \citenamefont {Toyooka}, \citenamefont {Shimosato},
  \citenamefont {Yagi}, \citenamefont {Takahashi}, \citenamefont {Okochi},
  \citenamefont {Okuda}, \citenamefont {Matoba}, \citenamefont {Sharov},
  \citenamefont {Ko},\ and\ \citenamefont {Niwa}}]{Masui2007}%
  \BibitemOpen
  \bibfield  {author} {\bibinfo {author} {\bibfnamefont {S.}~\bibnamefont
  {Masui}}, \bibinfo {author} {\bibfnamefont {Y.}~\bibnamefont {Nakatake}},
  \bibinfo {author} {\bibfnamefont {Y.}~\bibnamefont {Toyooka}}, \bibinfo
  {author} {\bibfnamefont {D.}~\bibnamefont {Shimosato}}, \bibinfo {author}
  {\bibfnamefont {R.}~\bibnamefont {Yagi}}, \bibinfo {author} {\bibfnamefont
  {K.}~\bibnamefont {Takahashi}}, \bibinfo {author} {\bibfnamefont
  {H.}~\bibnamefont {Okochi}}, \bibinfo {author} {\bibfnamefont
  {A.}~\bibnamefont {Okuda}}, \bibinfo {author} {\bibfnamefont
  {R.}~\bibnamefont {Matoba}}, \bibinfo {author} {\bibfnamefont {A.~A.}\
  \bibnamefont {Sharov}}, \bibinfo {author} {\bibfnamefont {M.~S.~H.}\
  \bibnamefont {Ko}},\ and\ \bibinfo {author} {\bibfnamefont {H.}~\bibnamefont
  {Niwa}},\ }\href@noop {} {\bibfield  {journal} {\bibinfo  {journal} {Nature
  Cell Biology}\ }\textbf {\bibinfo {volume} {9}},\ \bibinfo {pages} {625}
  (\bibinfo {year} {2007})}\BibitemShut {NoStop}%
\bibitem [{\citenamefont {Mitsui}\ and\ \citenamefont
  {et~al.}(2003)}]{Mitsui2003Nanog}%
  \BibitemOpen
  \bibfield  {author} {\bibinfo {author} {\bibfnamefont {K.}~\bibnamefont
  {Mitsui}}\ and\ \bibinfo {author} {\bibnamefont {et~al.}},\ }\href
  {https://doi.org/10.1016/S0092-8674(03)00393-3} {\bibfield  {journal}
  {\bibinfo  {journal} {Cell}\ }\textbf {\bibinfo {volume} {113}},\ \bibinfo
  {pages} {631} (\bibinfo {year} {2003})}\BibitemShut {NoStop}%
\bibitem [{\citenamefont {Karwacki-Neisius}\ \emph {et~al.}(2013)\citenamefont
  {Karwacki-Neisius}, \citenamefont {Göke}, \citenamefont {Osorno},
  \citenamefont {Halbritter}, \citenamefont {Ng}, \citenamefont {Weiße},
  \citenamefont {Wong}, \citenamefont {Gagliardi}, \citenamefont {Mullin},
  \citenamefont {Festuccia}, \citenamefont {Colby}, \citenamefont {Tomlinson},
  \citenamefont {Ng},\ and\ \citenamefont {Chambers}}]{KarwackiNeisius2013}%
  \BibitemOpen
  \bibfield  {author} {\bibinfo {author} {\bibfnamefont {V.}~\bibnamefont
  {Karwacki-Neisius}}, \bibinfo {author} {\bibfnamefont {J.}~\bibnamefont
  {Göke}}, \bibinfo {author} {\bibfnamefont {R.}~\bibnamefont {Osorno}},
  \bibinfo {author} {\bibfnamefont {F.}~\bibnamefont {Halbritter}}, \bibinfo
  {author} {\bibfnamefont {J.~H.}\ \bibnamefont {Ng}}, \bibinfo {author}
  {\bibfnamefont {A.~Y.}\ \bibnamefont {Weiße}}, \bibinfo {author}
  {\bibfnamefont {F.~C.~K.}\ \bibnamefont {Wong}}, \bibinfo {author}
  {\bibfnamefont {A.}~\bibnamefont {Gagliardi}}, \bibinfo {author}
  {\bibfnamefont {N.~P.}\ \bibnamefont {Mullin}}, \bibinfo {author}
  {\bibfnamefont {N.}~\bibnamefont {Festuccia}}, \bibinfo {author}
  {\bibfnamefont {D.}~\bibnamefont {Colby}}, \bibinfo {author} {\bibfnamefont
  {S.~R.}\ \bibnamefont {Tomlinson}}, \bibinfo {author} {\bibfnamefont {H.-H.}\
  \bibnamefont {Ng}},\ and\ \bibinfo {author} {\bibfnamefont {I.}~\bibnamefont
  {Chambers}},\ }\href {https://doi.org/10.1016/j.stem.2013.04.023} {\bibfield
  {journal} {\bibinfo  {journal} {Cell Stem Cell}\ }\textbf {\bibinfo {volume}
  {12}},\ \bibinfo {pages} {531} (\bibinfo {year} {2013})}\BibitemShut
  {NoStop}%
\bibitem [{\citenamefont {Chambers}\ \emph {et~al.}(2007)\citenamefont
  {Chambers}, \citenamefont {Silva}, \citenamefont {Colby}, \citenamefont
  {Nichols}, \citenamefont {Nijmeijer}, \citenamefont {Robertson},
  \citenamefont {Vrana}, \citenamefont {Jones}, \citenamefont {Grotewold},\
  and\ \citenamefont {Smith}}]{Chambers2007}%
  \BibitemOpen
  \bibfield  {author} {\bibinfo {author} {\bibfnamefont {I.}~\bibnamefont
  {Chambers}}, \bibinfo {author} {\bibfnamefont {J.}~\bibnamefont {Silva}},
  \bibinfo {author} {\bibfnamefont {D.}~\bibnamefont {Colby}}, \bibinfo
  {author} {\bibfnamefont {J.}~\bibnamefont {Nichols}}, \bibinfo {author}
  {\bibfnamefont {B.}~\bibnamefont {Nijmeijer}}, \bibinfo {author}
  {\bibfnamefont {M.}~\bibnamefont {Robertson}}, \bibinfo {author}
  {\bibfnamefont {J.}~\bibnamefont {Vrana}}, \bibinfo {author} {\bibfnamefont
  {K.}~\bibnamefont {Jones}}, \bibinfo {author} {\bibfnamefont
  {L.}~\bibnamefont {Grotewold}},\ and\ \bibinfo {author} {\bibfnamefont
  {A.}~\bibnamefont {Smith}},\ }\href@noop {} {\bibfield  {journal} {\bibinfo
  {journal} {Nature}\ }\textbf {\bibinfo {volume} {450}},\ \bibinfo {pages}
  {1230} (\bibinfo {year} {2007})}\BibitemShut {NoStop}%
\bibitem [{\citenamefont {Mullin}\ \emph {et~al.}(2008)\citenamefont {Mullin},
  \citenamefont {Yates}, \citenamefont {Rowe}, \citenamefont {Nijmeijer},
  \citenamefont {Colby}, \citenamefont {Barlow}, \citenamefont {Walkinshaw},\
  and\ \citenamefont {Chambers}}]{Mullin2008}%
  \BibitemOpen
  \bibfield  {author} {\bibinfo {author} {\bibfnamefont {N.}~\bibnamefont
  {Mullin}}, \bibinfo {author} {\bibfnamefont {A.}~\bibnamefont {Yates}},
  \bibinfo {author} {\bibfnamefont {A.}~\bibnamefont {Rowe}}, \bibinfo {author}
  {\bibfnamefont {B.}~\bibnamefont {Nijmeijer}}, \bibinfo {author}
  {\bibfnamefont {D.}~\bibnamefont {Colby}}, \bibinfo {author} {\bibfnamefont
  {P.}~\bibnamefont {Barlow}}, \bibinfo {author} {\bibfnamefont
  {M.}~\bibnamefont {Walkinshaw}},\ and\ \bibinfo {author} {\bibfnamefont
  {I.}~\bibnamefont {Chambers}},\ }\href {https://doi.org/10.1042/BJ20080134}
  {\bibfield  {journal} {\bibinfo  {journal} {Biochemical Journal}\ }\textbf
  {\bibinfo {volume} {411}},\ \bibinfo {pages} {227} (\bibinfo {year}
  {2008})}\BibitemShut {NoStop}%
\bibitem [{\citenamefont {Wang}\ \emph {et~al.}(2008)\citenamefont {Wang},
  \citenamefont {Levasseur},\ and\ \citenamefont {Orkin}}]{Wang2008}%
  \BibitemOpen
  \bibfield  {author} {\bibinfo {author} {\bibfnamefont {J.}~\bibnamefont
  {Wang}}, \bibinfo {author} {\bibfnamefont {D.~N.}\ \bibnamefont
  {Levasseur}},\ and\ \bibinfo {author} {\bibfnamefont {S.~H.}\ \bibnamefont
  {Orkin}},\ }\href {https://doi.org/10.1073/pnas.0802288105} {\bibfield
  {journal} {\bibinfo  {journal} {Proceedings of the National Academy of
  Sciences of the United States of America}\ }\textbf {\bibinfo {volume}
  {105}},\ \bibinfo {pages} {6326} (\bibinfo {year} {2008})}\BibitemShut
  {NoStop}%
\bibitem [{\citenamefont {Choi}\ \emph {et~al.}(2022)\citenamefont {Choi},
  \citenamefont {Quan}, \citenamefont {Qi}, \citenamefont {Lee}, \citenamefont
  {Tsoi}, \citenamefont {Zahabiyon}, \citenamefont {Bajic}, \citenamefont {Hu},
  \citenamefont {Prasad}, \citenamefont {Liao}, \citenamefont {Li},
  \citenamefont {Ferreon},\ and\ \citenamefont {Ferreon}}]{Choi2022}%
  \BibitemOpen
  \bibfield  {author} {\bibinfo {author} {\bibfnamefont {K.-J.}\ \bibnamefont
  {Choi}}, \bibinfo {author} {\bibfnamefont {M.~D.}\ \bibnamefont {Quan}},
  \bibinfo {author} {\bibfnamefont {C.}~\bibnamefont {Qi}}, \bibinfo {author}
  {\bibfnamefont {J.-H.}\ \bibnamefont {Lee}}, \bibinfo {author} {\bibfnamefont
  {P.~S.}\ \bibnamefont {Tsoi}}, \bibinfo {author} {\bibfnamefont
  {M.}~\bibnamefont {Zahabiyon}}, \bibinfo {author} {\bibfnamefont
  {A.}~\bibnamefont {Bajic}}, \bibinfo {author} {\bibfnamefont
  {L.}~\bibnamefont {Hu}}, \bibinfo {author} {\bibfnamefont {B.~V.~V.}\
  \bibnamefont {Prasad}}, \bibinfo {author} {\bibfnamefont {S.-C.~J.}\
  \bibnamefont {Liao}}, \bibinfo {author} {\bibfnamefont {W.}~\bibnamefont
  {Li}}, \bibinfo {author} {\bibfnamefont {A.~C.~M.}\ \bibnamefont {Ferreon}},\
  and\ \bibinfo {author} {\bibfnamefont {J.~C.}\ \bibnamefont {Ferreon}},\
  }\href@noop {} {\bibfield  {journal} {\bibinfo  {journal} {Nature Cell
  Biology}\ }\textbf {\bibinfo {volume} {24}},\ \bibinfo {pages} {737}
  (\bibinfo {year} {2022})}\BibitemShut {NoStop}%
\bibitem [{\citenamefont {Boija}\ \emph {et~al.}(2018)\citenamefont {Boija},
  \citenamefont {Klein}, \citenamefont {Sabari}, \citenamefont {Dall'Agnese},
  \citenamefont {Coffey}, \citenamefont {Zamudio}, \citenamefont {Li},
  \citenamefont {Shrinivas}, \citenamefont {Manteiga}, \citenamefont {Hannett},
  \citenamefont {Abraham}, \citenamefont {Afeyan}, \citenamefont {Guo},
  \citenamefont {Rimel}, \citenamefont {Fant}, \citenamefont {Schuijers},
  \citenamefont {Lee}, \citenamefont {Taatjes},\ and\ \citenamefont
  {Young}}]{Boija2018}%
  \BibitemOpen
  \bibfield  {author} {\bibinfo {author} {\bibfnamefont {A.}~\bibnamefont
  {Boija}}, \bibinfo {author} {\bibfnamefont {I.~A.}\ \bibnamefont {Klein}},
  \bibinfo {author} {\bibfnamefont {B.~R.}\ \bibnamefont {Sabari}}, \bibinfo
  {author} {\bibfnamefont {A.}~\bibnamefont {Dall'Agnese}}, \bibinfo {author}
  {\bibfnamefont {E.~L.}\ \bibnamefont {Coffey}}, \bibinfo {author}
  {\bibfnamefont {A.~V.}\ \bibnamefont {Zamudio}}, \bibinfo {author}
  {\bibfnamefont {C.~H.}\ \bibnamefont {Li}}, \bibinfo {author} {\bibfnamefont
  {K.}~\bibnamefont {Shrinivas}}, \bibinfo {author} {\bibfnamefont {J.~C.}\
  \bibnamefont {Manteiga}}, \bibinfo {author} {\bibfnamefont {N.~M.}\
  \bibnamefont {Hannett}}, \bibinfo {author} {\bibfnamefont {B.~J.}\
  \bibnamefont {Abraham}}, \bibinfo {author} {\bibfnamefont {L.~K.}\
  \bibnamefont {Afeyan}}, \bibinfo {author} {\bibfnamefont {Y.~E.}\
  \bibnamefont {Guo}}, \bibinfo {author} {\bibfnamefont {J.~K.}\ \bibnamefont
  {Rimel}}, \bibinfo {author} {\bibfnamefont {C.~B.}\ \bibnamefont {Fant}},
  \bibinfo {author} {\bibfnamefont {J.}~\bibnamefont {Schuijers}}, \bibinfo
  {author} {\bibfnamefont {T.~I.}\ \bibnamefont {Lee}}, \bibinfo {author}
  {\bibfnamefont {D.~J.}\ \bibnamefont {Taatjes}},\ and\ \bibinfo {author}
  {\bibfnamefont {R.~A.}\ \bibnamefont {Young}},\ }\href@noop {} {\bibfield
  {journal} {\bibinfo  {journal} {Cell}\ }\textbf {\bibinfo {volume} {175}},\
  \bibinfo {pages} {1842} (\bibinfo {year} {2018})}\BibitemShut {NoStop}%
\bibitem [{\citenamefont {McSwiggen}\ \emph {et~al.}(2019)\citenamefont
  {McSwiggen}, \citenamefont {Mir}, \citenamefont {Darzacq},\ and\
  \citenamefont {Tjian}}]{McSwiggen2019}%
  \BibitemOpen
  \bibfield  {author} {\bibinfo {author} {\bibfnamefont {D.~T.}\ \bibnamefont
  {McSwiggen}}, \bibinfo {author} {\bibfnamefont {M.}~\bibnamefont {Mir}},
  \bibinfo {author} {\bibfnamefont {X.}~\bibnamefont {Darzacq}},\ and\ \bibinfo
  {author} {\bibfnamefont {R.}~\bibnamefont {Tjian}},\ }\href@noop {}
  {\bibfield  {journal} {\bibinfo  {journal} {Genes \& development}\ }\textbf
  {\bibinfo {volume} {33}},\ \bibinfo {pages} {1619} (\bibinfo {year}
  {2019})}\BibitemShut {NoStop}%
\bibitem [{\citenamefont {Mullin}\ \emph {et~al.}(2017)\citenamefont {Mullin},
  \citenamefont {Gagliardi}, \citenamefont {Khoa}, \citenamefont {Colby},
  \citenamefont {Hall-Ponsele}, \citenamefont {Rowe},\ and\ \citenamefont
  {Chambers}}]{Mullin2017}%
  \BibitemOpen
  \bibfield  {author} {\bibinfo {author} {\bibfnamefont {N.~P.}\ \bibnamefont
  {Mullin}}, \bibinfo {author} {\bibfnamefont {A.}~\bibnamefont {Gagliardi}},
  \bibinfo {author} {\bibfnamefont {L.~T.~P.}\ \bibnamefont {Khoa}}, \bibinfo
  {author} {\bibfnamefont {D.}~\bibnamefont {Colby}}, \bibinfo {author}
  {\bibfnamefont {E.}~\bibnamefont {Hall-Ponsele}}, \bibinfo {author}
  {\bibfnamefont {A.~J.}\ \bibnamefont {Rowe}},\ and\ \bibinfo {author}
  {\bibfnamefont {I.}~\bibnamefont {Chambers}},\ }\href
  {https://doi.org/https://doi.org/10.1016/j.jmb.2016.12.001} {\bibfield
  {journal} {\bibinfo  {journal} {Journal of Molecular Biology}\ }\textbf
  {\bibinfo {volume} {429}},\ \bibinfo {pages} {1544} (\bibinfo {year}
  {2017})}\BibitemShut {NoStop}%
\bibitem [{\citenamefont {Lopes~Novo}\ \emph {et~al.}(2016)\citenamefont
  {Lopes~Novo}, \citenamefont {Tang}, \citenamefont {Ahmed}, \citenamefont
  {Djuric}, \citenamefont {Fussner}, \citenamefont {Mullin}, \citenamefont
  {Morgan}, \citenamefont {Hayre}, \citenamefont {Sienerth}, \citenamefont
  {Elderkin}, \citenamefont {Nishinakamura}, \citenamefont {Chambers},
  \citenamefont {Ellis}, \citenamefont {Bazett-Jones},\ and\ \citenamefont
  {Rugg-Gunn}}]{Novo2016Nanog}%
  \BibitemOpen
  \bibfield  {author} {\bibinfo {author} {\bibfnamefont {C.}~\bibnamefont
  {Lopes~Novo}}, \bibinfo {author} {\bibfnamefont {C.}~\bibnamefont {Tang}},
  \bibinfo {author} {\bibfnamefont {K.}~\bibnamefont {Ahmed}}, \bibinfo
  {author} {\bibfnamefont {U.}~\bibnamefont {Djuric}}, \bibinfo {author}
  {\bibfnamefont {E.}~\bibnamefont {Fussner}}, \bibinfo {author} {\bibfnamefont
  {N.~P.}\ \bibnamefont {Mullin}}, \bibinfo {author} {\bibfnamefont {N.~P.}\
  \bibnamefont {Morgan}}, \bibinfo {author} {\bibfnamefont {J.}~\bibnamefont
  {Hayre}}, \bibinfo {author} {\bibfnamefont {A.~R.}\ \bibnamefont {Sienerth}},
  \bibinfo {author} {\bibfnamefont {S.}~\bibnamefont {Elderkin}}, \bibinfo
  {author} {\bibfnamefont {R.}~\bibnamefont {Nishinakamura}}, \bibinfo {author}
  {\bibfnamefont {I.}~\bibnamefont {Chambers}}, \bibinfo {author}
  {\bibfnamefont {J.}~\bibnamefont {Ellis}}, \bibinfo {author} {\bibfnamefont
  {D.~P.}\ \bibnamefont {Bazett-Jones}},\ and\ \bibinfo {author} {\bibfnamefont
  {P.~J.}\ \bibnamefont {Rugg-Gunn}},\ }\href
  {https://doi.org/10.1101/gad.275685.115} {\bibfield  {journal} {\bibinfo
  {journal} {Genes \& Development}\ }\textbf {\bibinfo {volume} {30}},\
  \bibinfo {pages} {1101} (\bibinfo {year} {2016})}\BibitemShut {NoStop}%
\bibitem [{\citenamefont {Lopes~Novo}\ \emph {et~al.}(2018)\citenamefont
  {Lopes~Novo}, \citenamefont {Javierre}, \citenamefont {Cairns}, \citenamefont
  {Segonds-Pichon}, \citenamefont {Wingett}, \citenamefont {Freire-Pritchett},
  \citenamefont {Furlan-Magaril}, \citenamefont {Schoenfelder}, \citenamefont
  {Fraser},\ and\ \citenamefont {Rugg-Gunn}}]{Novo2018LongRange}%
  \BibitemOpen
  \bibfield  {author} {\bibinfo {author} {\bibfnamefont {C.}~\bibnamefont
  {Lopes~Novo}}, \bibinfo {author} {\bibfnamefont {B.-M.}\ \bibnamefont
  {Javierre}}, \bibinfo {author} {\bibfnamefont {J.}~\bibnamefont {Cairns}},
  \bibinfo {author} {\bibfnamefont {A.}~\bibnamefont {Segonds-Pichon}},
  \bibinfo {author} {\bibfnamefont {S.~W.}\ \bibnamefont {Wingett}}, \bibinfo
  {author} {\bibfnamefont {P.}~\bibnamefont {Freire-Pritchett}}, \bibinfo
  {author} {\bibfnamefont {M.}~\bibnamefont {Furlan-Magaril}}, \bibinfo
  {author} {\bibfnamefont {S.}~\bibnamefont {Schoenfelder}}, \bibinfo {author}
  {\bibfnamefont {P.}~\bibnamefont {Fraser}},\ and\ \bibinfo {author}
  {\bibfnamefont {P.~J.}\ \bibnamefont {Rugg-Gunn}},\ }\href
  {https://doi.org/10.1016/j.celrep.2018.02.040} {\bibfield  {journal}
  {\bibinfo  {journal} {Cell Reports}\ }\textbf {\bibinfo {volume} {22}},\
  \bibinfo {pages} {2615} (\bibinfo {year} {2018})}\BibitemShut {NoStop}%
\bibitem [{\citenamefont {de~Wit}\ \emph {et~al.}(2013)\citenamefont {de~Wit},
  \citenamefont {Bouwman}, \citenamefont {Zhu}, \citenamefont {Klous},
  \citenamefont {Splinter}, \citenamefont {Verstegen}, \citenamefont {Krijger},
  \citenamefont {Festuccia}, \citenamefont {Nora}, \citenamefont {Welling},
  \citenamefont {Heard}, \citenamefont {Geijsen}, \citenamefont {Poot},
  \citenamefont {Chambers},\ and\ \citenamefont
  {de~Laat}}]{deWit2013Pluripotent}%
  \BibitemOpen
  \bibfield  {author} {\bibinfo {author} {\bibfnamefont {E.}~\bibnamefont
  {de~Wit}}, \bibinfo {author} {\bibfnamefont {B.~A.~M.}\ \bibnamefont
  {Bouwman}}, \bibinfo {author} {\bibfnamefont {Y.}~\bibnamefont {Zhu}},
  \bibinfo {author} {\bibfnamefont {P.}~\bibnamefont {Klous}}, \bibinfo
  {author} {\bibfnamefont {E.}~\bibnamefont {Splinter}}, \bibinfo {author}
  {\bibfnamefont {M.~J. A.~M.}\ \bibnamefont {Verstegen}}, \bibinfo {author}
  {\bibfnamefont {P.~H.~L.}\ \bibnamefont {Krijger}}, \bibinfo {author}
  {\bibfnamefont {N.}~\bibnamefont {Festuccia}}, \bibinfo {author}
  {\bibfnamefont {E.~P.}\ \bibnamefont {Nora}}, \bibinfo {author}
  {\bibfnamefont {M.}~\bibnamefont {Welling}}, \bibinfo {author} {\bibfnamefont
  {E.}~\bibnamefont {Heard}}, \bibinfo {author} {\bibfnamefont
  {N.}~\bibnamefont {Geijsen}}, \bibinfo {author} {\bibfnamefont {R.~A.}\
  \bibnamefont {Poot}}, \bibinfo {author} {\bibfnamefont {I.}~\bibnamefont
  {Chambers}},\ and\ \bibinfo {author} {\bibfnamefont {W.}~\bibnamefont
  {de~Laat}},\ }\href {https://doi.org/10.1038/nature12420} {\bibfield
  {journal} {\bibinfo  {journal} {Nature}\ }\textbf {\bibinfo {volume} {501}},\
  \bibinfo {pages} {227} (\bibinfo {year} {2013})}\BibitemShut {NoStop}%
\bibitem [{\citenamefont {Gagliardi}\ \emph {et~al.}(2013)\citenamefont
  {Gagliardi}, \citenamefont {Mullin}, \citenamefont {Ying~Tan}, \citenamefont
  {Colby}, \citenamefont {Kousa}, \citenamefont {Halbritter}, \citenamefont
  {Weiss}, \citenamefont {Felker}, \citenamefont {Bezstarosti}, \citenamefont
  {Favaro}, \citenamefont {Demmers}, \citenamefont {Nicolis}, \citenamefont
  {Tomlinson}, \citenamefont {Poot},\ and\ \citenamefont
  {Chambers}}]{Gagliardi2013}%
  \BibitemOpen
  \bibfield  {author} {\bibinfo {author} {\bibfnamefont {A.}~\bibnamefont
  {Gagliardi}}, \bibinfo {author} {\bibfnamefont {N.~P.}\ \bibnamefont
  {Mullin}}, \bibinfo {author} {\bibfnamefont {Z.}~\bibnamefont {Ying~Tan}},
  \bibinfo {author} {\bibfnamefont {D.}~\bibnamefont {Colby}}, \bibinfo
  {author} {\bibfnamefont {A.~I.}\ \bibnamefont {Kousa}}, \bibinfo {author}
  {\bibfnamefont {F.}~\bibnamefont {Halbritter}}, \bibinfo {author}
  {\bibfnamefont {J.~T.}\ \bibnamefont {Weiss}}, \bibinfo {author}
  {\bibfnamefont {A.}~\bibnamefont {Felker}}, \bibinfo {author} {\bibfnamefont
  {K.}~\bibnamefont {Bezstarosti}}, \bibinfo {author} {\bibfnamefont
  {R.}~\bibnamefont {Favaro}}, \bibinfo {author} {\bibfnamefont
  {J.}~\bibnamefont {Demmers}}, \bibinfo {author} {\bibfnamefont {S.~K.}\
  \bibnamefont {Nicolis}}, \bibinfo {author} {\bibfnamefont {S.~R.}\
  \bibnamefont {Tomlinson}}, \bibinfo {author} {\bibfnamefont {R.~A.}\
  \bibnamefont {Poot}},\ and\ \bibinfo {author} {\bibfnamefont
  {I.}~\bibnamefont {Chambers}},\ }\href
  {https://doi.org/https://doi.org/10.1038/emboj.2013.161} {\bibfield
  {journal} {\bibinfo  {journal} {The EMBO Journal}\ }\textbf {\bibinfo
  {volume} {32}},\ \bibinfo {pages} {2231} (\bibinfo {year}
  {2013})}\BibitemShut {NoStop}%
\bibitem [{\citenamefont {Jawerth}\ \emph {et~al.}(2020)\citenamefont
  {Jawerth}, \citenamefont {Fischer-Friedrich}, \citenamefont {Saha},
  \citenamefont {Wang}, \citenamefont {Franzmann}, \citenamefont {Zhang},
  \citenamefont {Sachweh}, \citenamefont {Ruer}, \citenamefont {Ijavi},
  \citenamefont {Saha}, \citenamefont {Mahamid}, \citenamefont {Hyman},\ and\
  \citenamefont {Jülicher}}]{Jawerth2020}%
  \BibitemOpen
  \bibfield  {author} {\bibinfo {author} {\bibfnamefont {L.}~\bibnamefont
  {Jawerth}}, \bibinfo {author} {\bibfnamefont {E.}~\bibnamefont
  {Fischer-Friedrich}}, \bibinfo {author} {\bibfnamefont {S.}~\bibnamefont
  {Saha}}, \bibinfo {author} {\bibfnamefont {J.}~\bibnamefont {Wang}}, \bibinfo
  {author} {\bibfnamefont {T.}~\bibnamefont {Franzmann}}, \bibinfo {author}
  {\bibfnamefont {X.}~\bibnamefont {Zhang}}, \bibinfo {author} {\bibfnamefont
  {J.}~\bibnamefont {Sachweh}}, \bibinfo {author} {\bibfnamefont
  {M.}~\bibnamefont {Ruer}}, \bibinfo {author} {\bibfnamefont {M.}~\bibnamefont
  {Ijavi}}, \bibinfo {author} {\bibfnamefont {S.}~\bibnamefont {Saha}},
  \bibinfo {author} {\bibfnamefont {J.}~\bibnamefont {Mahamid}}, \bibinfo
  {author} {\bibfnamefont {A.~A.}\ \bibnamefont {Hyman}},\ and\ \bibinfo
  {author} {\bibfnamefont {F.}~\bibnamefont {Jülicher}},\ }\href
  {https://doi.org/10.1126/science.aaw4951} {\bibfield  {journal} {\bibinfo
  {journal} {Science}\ }\textbf {\bibinfo {volume} {370}},\ \bibinfo {pages}
  {1317} (\bibinfo {year} {2020})}\BibitemShut {NoStop}%
\bibitem [{\citenamefont {Navarro}\ \emph {et~al.}(2012)\citenamefont
  {Navarro}, \citenamefont {Festuccia}, \citenamefont {Colby}, \citenamefont
  {Gagliardi}, \citenamefont {Mullin}, \citenamefont {Zhang}, \citenamefont
  {Karwacki-Neisius}, \citenamefont {Osorno}, \citenamefont {Kelly},
  \citenamefont {Robertson},\ and\ \citenamefont
  {Chambers}}]{Navarro2012NanogAutorepression}%
  \BibitemOpen
  \bibfield  {author} {\bibinfo {author} {\bibfnamefont {P.}~\bibnamefont
  {Navarro}}, \bibinfo {author} {\bibfnamefont {N.}~\bibnamefont {Festuccia}},
  \bibinfo {author} {\bibfnamefont {D.}~\bibnamefont {Colby}}, \bibinfo
  {author} {\bibfnamefont {A.}~\bibnamefont {Gagliardi}}, \bibinfo {author}
  {\bibfnamefont {N.~P.}\ \bibnamefont {Mullin}}, \bibinfo {author}
  {\bibfnamefont {W.}~\bibnamefont {Zhang}}, \bibinfo {author} {\bibfnamefont
  {V.}~\bibnamefont {Karwacki-Neisius}}, \bibinfo {author} {\bibfnamefont
  {R.}~\bibnamefont {Osorno}}, \bibinfo {author} {\bibfnamefont
  {D.}~\bibnamefont {Kelly}}, \bibinfo {author} {\bibfnamefont
  {M.}~\bibnamefont {Robertson}},\ and\ \bibinfo {author} {\bibfnamefont
  {I.}~\bibnamefont {Chambers}},\ }\href
  {https://doi.org/10.1038/emboj.2012.321} {\bibfield  {journal} {\bibinfo
  {journal} {The EMBO Journal}\ }\textbf {\bibinfo {volume} {31}},\ \bibinfo
  {pages} {4547} (\bibinfo {year} {2012})}\BibitemShut {NoStop}%
\bibitem [{\citenamefont {Jauch}\ \emph {et~al.}(2008)\citenamefont {Jauch},
  \citenamefont {Ng}, \citenamefont {Saikatendu}, \citenamefont {Stevens},\
  and\ \citenamefont {Kolatkar}}]{Jauch2008}%
  \BibitemOpen
  \bibfield  {author} {\bibinfo {author} {\bibfnamefont {R.}~\bibnamefont
  {Jauch}}, \bibinfo {author} {\bibfnamefont {C.~K.~L.}\ \bibnamefont {Ng}},
  \bibinfo {author} {\bibfnamefont {K.~S.}\ \bibnamefont {Saikatendu}},
  \bibinfo {author} {\bibfnamefont {R.~C.}\ \bibnamefont {Stevens}},\ and\
  \bibinfo {author} {\bibfnamefont {P.~R.}\ \bibnamefont {Kolatkar}},\ }\href
  {https://doi.org/https://doi.org/10.1016/j.jmb.2007.11.091} {\bibfield
  {journal} {\bibinfo  {journal} {Journal of Molecular Biology}\ }\textbf
  {\bibinfo {volume} {376}},\ \bibinfo {pages} {758} (\bibinfo {year}
  {2008})}\BibitemShut {NoStop}%
\bibitem [{\citenamefont {Michieletto}\ and\ \citenamefont
  {Marenda}(2022)}]{Michieletto2022}%
  \BibitemOpen
  \bibfield  {author} {\bibinfo {author} {\bibfnamefont {D.}~\bibnamefont
  {Michieletto}}\ and\ \bibinfo {author} {\bibfnamefont {M.}~\bibnamefont
  {Marenda}},\ }\href {https://doi.org/10.1021/jacsau.2c00055} {\bibfield
  {journal} {\bibinfo  {journal} {JACS Au}\ }\textbf {\bibinfo {volume} {2}},\
  \bibinfo {pages} {1506} (\bibinfo {year} {2022})}\BibitemShut {NoStop}%
\bibitem [{\citenamefont {Cicuta}\ and\ \citenamefont
  {Donald}(2007)}]{Cicuta2007}%
  \BibitemOpen
  \bibfield  {author} {\bibinfo {author} {\bibfnamefont {P.}~\bibnamefont
  {Cicuta}}\ and\ \bibinfo {author} {\bibfnamefont {A.~M.}\ \bibnamefont
  {Donald}},\ }\href {https://doi.org/10.1039/b706004c} {\bibfield  {journal}
  {\bibinfo  {journal} {Soft Matter}\ }\textbf {\bibinfo {volume} {3}},\
  \bibinfo {pages} {1449} (\bibinfo {year} {2007})}\BibitemShut {NoStop}%
\bibitem [{\citenamefont {Mason}(2000)}]{Mason2000}%
  \BibitemOpen
  \bibfield  {author} {\bibinfo {author} {\bibfnamefont {T.~G.}\ \bibnamefont
  {Mason}},\ }\href {https://doi.org/10.1007/s003970000094} {\bibfield
  {journal} {\bibinfo  {journal} {Rheologica Acta}\ }\textbf {\bibinfo {volume}
  {39}},\ \bibinfo {pages} {371} (\bibinfo {year} {2000})}\BibitemShut
  {NoStop}%
\bibitem [{\citenamefont {Mason}\ and\ \citenamefont
  {Weitz}(1995)}]{Mason1995}%
  \BibitemOpen
  \bibfield  {author} {\bibinfo {author} {\bibfnamefont {T.~G.}\ \bibnamefont
  {Mason}}\ and\ \bibinfo {author} {\bibfnamefont {D.~A.}\ \bibnamefont
  {Weitz}},\ }\href {https://doi.org/10.1103/PhysRevLett.74.1250} {\bibfield
  {journal} {\bibinfo  {journal} {Physical Review Letters}\ }\textbf {\bibinfo
  {volume} {74}},\ \bibinfo {pages} {1250} (\bibinfo {year}
  {1995})}\BibitemShut {NoStop}%
\bibitem [{\citenamefont {Brizioli}\ \emph {et~al.}(2025)\citenamefont
  {Brizioli}, \citenamefont {Escobedo-S{'a}nchez}, \citenamefont {McCall},
  \citenamefont {Roichman}, \citenamefont {Beck~Adiels}, \citenamefont
  {Volpe},\ and\ \citenamefont {Cerbino}}]{Brizioli2025OFM}%
  \BibitemOpen
  \bibfield  {author} {\bibinfo {author} {\bibfnamefont {M.}~\bibnamefont
  {Brizioli}}, \bibinfo {author} {\bibfnamefont {M.~A.}\ \bibnamefont
  {Escobedo-S{'a}nchez}}, \bibinfo {author} {\bibfnamefont {P.~M.}\
  \bibnamefont {McCall}}, \bibinfo {author} {\bibfnamefont {Y.}~\bibnamefont
  {Roichman}}, \bibinfo {author} {\bibfnamefont {C.}~\bibnamefont
  {Beck~Adiels}}, \bibinfo {author} {\bibfnamefont {G.}~\bibnamefont {Volpe}},\
  and\ \bibinfo {author} {\bibfnamefont {R.}~\bibnamefont {Cerbino}},\ }\href
  {https://doi.org/10.1039/d4sm01390e} {\bibfield  {journal} {\bibinfo
  {journal} {Soft Matter}\ }\textbf {\bibinfo {volume} {21}},\ \bibinfo {pages}
  {1234} (\bibinfo {year} {2025})}\BibitemShut {NoStop}%
\bibitem [{\citenamefont {Joseph}\ \emph {et~al.}(2021)\citenamefont {Joseph},
  \citenamefont {Reinhardt}, \citenamefont {Aguirre}, \citenamefont {Chew},
  \citenamefont {Russell}, \citenamefont {Espinosa}, \citenamefont {Garaizar},\
  and\ \citenamefont {Collepardo-Guevara}}]{Mpipi2021}%
  \BibitemOpen
  \bibfield  {author} {\bibinfo {author} {\bibfnamefont {J.~A.}\ \bibnamefont
  {Joseph}}, \bibinfo {author} {\bibfnamefont {A.}~\bibnamefont {Reinhardt}},
  \bibinfo {author} {\bibfnamefont {A.}~\bibnamefont {Aguirre}}, \bibinfo
  {author} {\bibfnamefont {P.~Y.}\ \bibnamefont {Chew}}, \bibinfo {author}
  {\bibfnamefont {K.~O.}\ \bibnamefont {Russell}}, \bibinfo {author}
  {\bibfnamefont {J.~R.}\ \bibnamefont {Espinosa}}, \bibinfo {author}
  {\bibfnamefont {A.}~\bibnamefont {Garaizar}},\ and\ \bibinfo {author}
  {\bibfnamefont {R.}~\bibnamefont {Collepardo-Guevara}},\ }\href@noop {}
  {\bibfield  {journal} {\bibinfo  {journal} {Nature Computational Science}\
  }\textbf {\bibinfo {volume} {1}},\ \bibinfo {pages} {732} (\bibinfo {year}
  {2021})}\BibitemShut {NoStop}%
\bibitem [{\citenamefont {Thompson}\ \emph {et~al.}(2022)\citenamefont
  {Thompson}, \citenamefont {Aktulga}, \citenamefont {Berger}, \citenamefont
  {Bolintineanu}, \citenamefont {Brown}, \citenamefont {Crozier}, \citenamefont
  {in~'t Veld}, \citenamefont {Kohlmeyer}, \citenamefont {Moore}, \citenamefont
  {Nguyen}, \citenamefont {Shan}, \citenamefont {Stevens}, \citenamefont
  {Tranchida}, \citenamefont {Trott},\ and\ \citenamefont
  {Plimpton}}]{LAMMPS2022}%
  \BibitemOpen
  \bibfield  {author} {\bibinfo {author} {\bibfnamefont {A.~P.}\ \bibnamefont
  {Thompson}}, \bibinfo {author} {\bibfnamefont {H.~M.}\ \bibnamefont
  {Aktulga}}, \bibinfo {author} {\bibfnamefont {R.}~\bibnamefont {Berger}},
  \bibinfo {author} {\bibfnamefont {D.~S.}\ \bibnamefont {Bolintineanu}},
  \bibinfo {author} {\bibfnamefont {W.~M.}\ \bibnamefont {Brown}}, \bibinfo
  {author} {\bibfnamefont {P.~S.}\ \bibnamefont {Crozier}}, \bibinfo {author}
  {\bibfnamefont {P.~J.}\ \bibnamefont {in~'t Veld}}, \bibinfo {author}
  {\bibfnamefont {A.}~\bibnamefont {Kohlmeyer}}, \bibinfo {author}
  {\bibfnamefont {S.~G.}\ \bibnamefont {Moore}}, \bibinfo {author}
  {\bibfnamefont {T.~D.}\ \bibnamefont {Nguyen}}, \bibinfo {author}
  {\bibfnamefont {R.}~\bibnamefont {Shan}}, \bibinfo {author} {\bibfnamefont
  {M.~J.}\ \bibnamefont {Stevens}}, \bibinfo {author} {\bibfnamefont
  {J.}~\bibnamefont {Tranchida}}, \bibinfo {author} {\bibfnamefont
  {C.}~\bibnamefont {Trott}},\ and\ \bibinfo {author} {\bibfnamefont {S.~J.}\
  \bibnamefont {Plimpton}},\ }\href {https://doi.org/10.1016/j.cpc.2021.108171}
  {\bibfield  {journal} {\bibinfo  {journal} {Comp. Phys. Comm.}\ }\textbf
  {\bibinfo {volume} {271}},\ \bibinfo {pages} {108171} (\bibinfo {year}
  {2022})}\BibitemShut {NoStop}%
\bibitem [{\citenamefont {Mizutani}\ \emph {et~al.}(2023)\citenamefont
  {Mizutani}, \citenamefont {Tan}, \citenamefont {Sugita},\ and\ \citenamefont
  {Takada}}]{Takada2023}%
  \BibitemOpen
  \bibfield  {author} {\bibinfo {author} {\bibfnamefont {A.}~\bibnamefont
  {Mizutani}}, \bibinfo {author} {\bibfnamefont {C.}~\bibnamefont {Tan}},
  \bibinfo {author} {\bibfnamefont {Y.}~\bibnamefont {Sugita}},\ and\ \bibinfo
  {author} {\bibfnamefont {S.}~\bibnamefont {Takada}},\ }\href@noop {}
  {\bibfield  {journal} {\bibinfo  {journal} {PLoS computational biology}\
  }\textbf {\bibinfo {volume} {19}},\ \bibinfo {pages} {e1011321} (\bibinfo
  {year} {2023})}\BibitemShut {NoStop}%
\bibitem [{\citenamefont {Hamley}(2007)}]{Hamley2007}%
  \BibitemOpen
  \bibfield  {author} {\bibinfo {author} {\bibfnamefont {I.~W.}\ \bibnamefont
  {Hamley}},\ }\href@noop {} {\emph {\bibinfo {title} {Introduction to Soft
  Matter: Synthetic and \linebreak Biological Self-Assembling Materials}}},\
  \bibinfo {edition} {2nd}\ ed.\ (\bibinfo  {publisher} {John Wiley \& Sons},\
  \bibinfo {year} {2007})\BibitemShut {NoStop}%
\bibitem [{\citenamefont {Zhu}\ \emph {et~al.}(2008)\citenamefont {Zhu},
  \citenamefont {Kundukad},\ and\ \citenamefont {{Van Der Maarel}}}]{Zhu2008}%
  \BibitemOpen
  \bibfield  {author} {\bibinfo {author} {\bibfnamefont {X.}~\bibnamefont
  {Zhu}}, \bibinfo {author} {\bibfnamefont {B.}~\bibnamefont {Kundukad}},\ and\
  \bibinfo {author} {\bibfnamefont {J.~R.}\ \bibnamefont {{Van Der Maarel}}},\
  }\href {https://doi.org/10.1063/1.3009249} {\bibfield  {journal} {\bibinfo
  {journal} {J. Chem. Phys.}\ }\textbf {\bibinfo {volume} {129}},\ \bibinfo
  {pages} {1} (\bibinfo {year} {2008})}\BibitemShut {NoStop}%
\bibitem [{\citenamefont {Muzzopappa}\ \emph {et~al.}(2021)\citenamefont
  {Muzzopappa}, \citenamefont {Hertzog},\ and\ \citenamefont
  {Erdel}}]{Muzzopappa2021}%
  \BibitemOpen
  \bibfield  {author} {\bibinfo {author} {\bibfnamefont {F.}~\bibnamefont
  {Muzzopappa}}, \bibinfo {author} {\bibfnamefont {M.}~\bibnamefont
  {Hertzog}},\ and\ \bibinfo {author} {\bibfnamefont {F.}~\bibnamefont
  {Erdel}},\ }\href {https://doi.org/10.1016/j.bpj.2021.02.027} {\bibfield
  {journal} {\bibinfo  {journal} {Biophysical Journal}\ }\textbf {\bibinfo
  {volume} {120}},\ \bibinfo {pages} {1288} (\bibinfo {year}
  {2021})}\BibitemShut {NoStop}%
\bibitem [{\citenamefont {Pulupa}\ \emph {et~al.}(2025)\citenamefont {Pulupa},
  \citenamefont {McArthur}, \citenamefont {Stathi}, \citenamefont {Wang},
  \citenamefont {Zazhytska}, \citenamefont {Pirozzolo}, \citenamefont {Nayar},
  \citenamefont {Shapiro},\ and\ \citenamefont {Lomvardas}}]{Pulupa2025}%
  \BibitemOpen
  \bibfield  {author} {\bibinfo {author} {\bibfnamefont {J.}~\bibnamefont
  {Pulupa}}, \bibinfo {author} {\bibfnamefont {N.~G.}\ \bibnamefont
  {McArthur}}, \bibinfo {author} {\bibfnamefont {O.}~\bibnamefont {Stathi}},
  \bibinfo {author} {\bibfnamefont {M.}~\bibnamefont {Wang}}, \bibinfo {author}
  {\bibfnamefont {M.}~\bibnamefont {Zazhytska}}, \bibinfo {author}
  {\bibfnamefont {I.~D.}\ \bibnamefont {Pirozzolo}}, \bibinfo {author}
  {\bibfnamefont {A.}~\bibnamefont {Nayar}}, \bibinfo {author} {\bibfnamefont
  {L.}~\bibnamefont {Shapiro}},\ and\ \bibinfo {author} {\bibfnamefont
  {S.}~\bibnamefont {Lomvardas}},\ }\href
  {https://doi.org/10.1038/s41586-025-09043-6} {\bibfield  {journal} {\bibinfo
  {journal} {Nature}\ }\textbf {\bibinfo {volume} {643}},\ \bibinfo {pages}
  {820} (\bibinfo {year} {2025})}\BibitemShut {NoStop}%
\bibitem [{\citenamefont {Mizutani}\ \emph {et~al.}(2025)\citenamefont
  {Mizutani}, \citenamefont {Tan}, \citenamefont {Sugita},\ and\ \citenamefont
  {Takada}}]{Takada2025}%
  \BibitemOpen
  \bibfield  {author} {\bibinfo {author} {\bibfnamefont {A.}~\bibnamefont
  {Mizutani}}, \bibinfo {author} {\bibfnamefont {C.}~\bibnamefont {Tan}},
  \bibinfo {author} {\bibfnamefont {Y.}~\bibnamefont {Sugita}},\ and\ \bibinfo
  {author} {\bibfnamefont {S.}~\bibnamefont {Takada}},\ }\href@noop {}
  {\bibfield  {journal} {\bibinfo  {journal} {Biophysical journal}\ }\textbf
  {\bibinfo {volume} {124}},\ \bibinfo {pages} {1587} (\bibinfo {year}
  {2025})}\BibitemShut {NoStop}%
\bibitem [{\citenamefont {Marenda}\ \emph {et~al.}(2024)\citenamefont
  {Marenda}, \citenamefont {Michieletto}, \citenamefont {Czapiewski},
  \citenamefont {Stocks}, \citenamefont {Winterbourne}, \citenamefont {Miles},
  \citenamefont {Flemming}, \citenamefont {Lazarova}, \citenamefont {Chiang},
  \citenamefont {Aitken}, \citenamefont {Grimes}, \citenamefont {Becher},
  \citenamefont {Cook}, \citenamefont {Marenduzzo}, \citenamefont {Nozawa},\
  and\ \citenamefont {Gilbert}}]{Marenda2024}%
  \BibitemOpen
  \bibfield  {author} {\bibinfo {author} {\bibfnamefont {M.}~\bibnamefont
  {Marenda}}, \bibinfo {author} {\bibfnamefont {D.}~\bibnamefont
  {Michieletto}}, \bibinfo {author} {\bibfnamefont {R.}~\bibnamefont
  {Czapiewski}}, \bibinfo {author} {\bibfnamefont {J.}~\bibnamefont {Stocks}},
  \bibinfo {author} {\bibfnamefont {S.~M.}\ \bibnamefont {Winterbourne}},
  \bibinfo {author} {\bibfnamefont {J.}~\bibnamefont {Miles}}, \bibinfo
  {author} {\bibfnamefont {O.~C.~A.}\ \bibnamefont {Flemming}}, \bibinfo
  {author} {\bibfnamefont {E.}~\bibnamefont {Lazarova}}, \bibinfo {author}
  {\bibfnamefont {M.}~\bibnamefont {Chiang}}, \bibinfo {author} {\bibfnamefont
  {S.}~\bibnamefont {Aitken}}, \bibinfo {author} {\bibfnamefont {G.~R.}\
  \bibnamefont {Grimes}}, \bibinfo {author} {\bibfnamefont {H.}~\bibnamefont
  {Becher}}, \bibinfo {author} {\bibfnamefont {A.}~\bibnamefont {Cook}},
  \bibinfo {author} {\bibfnamefont {D.}~\bibnamefont {Marenduzzo}}, \bibinfo
  {author} {\bibfnamefont {R.-S.}\ \bibnamefont {Nozawa}},\ and\ \bibinfo
  {author} {\bibfnamefont {N.}~\bibnamefont {Gilbert}},\ }\bibfield  {journal}
  {\bibinfo  {journal} {bioRxiv}\ }\href
  {https://doi.org/10.1101/2024.06.16.599208} {10.1101/2024.06.16.599208}
  (\bibinfo {year} {2024})\BibitemShut {NoStop}%
\bibitem [{\citenamefont {Li}\ \emph {et~al.}(2025)\citenamefont {Li},
  \citenamefont {Dalgleish}, \citenamefont {Lister}, \citenamefont {Maristany},
  \citenamefont {Huertas}, \citenamefont {Dopico-Fernandez}, \citenamefont
  {Hamley}, \citenamefont {Denny}, \citenamefont {Bloye}, \citenamefont
  {Zhang}, \citenamefont {Hentges}, \citenamefont {Doll}, \citenamefont {Wei},
  \citenamefont {Maresca}, \citenamefont {Dimitrova}, \citenamefont {Pytowski},
  \citenamefont {Tunnacliffe}, \citenamefont {Kassouf}, \citenamefont {Higgs},
  \citenamefont {de~Wit}, \citenamefont {Klose}, \citenamefont {Schermelleh},
  \citenamefont {Collepardo-Guevara}, \citenamefont {Milne},\ and\
  \citenamefont {Davies}}]{Li2025Chromatin}%
  \BibitemOpen
  \bibfield  {author} {\bibinfo {author} {\bibfnamefont {H.}~\bibnamefont
  {Li}}, \bibinfo {author} {\bibfnamefont {J.~L.}\ \bibnamefont {Dalgleish}},
  \bibinfo {author} {\bibfnamefont {G.}~\bibnamefont {Lister}}, \bibinfo
  {author} {\bibfnamefont {M.~J.}\ \bibnamefont {Maristany}}, \bibinfo {author}
  {\bibfnamefont {J.}~\bibnamefont {Huertas}}, \bibinfo {author} {\bibfnamefont
  {A.~M.}\ \bibnamefont {Dopico-Fernandez}}, \bibinfo {author} {\bibfnamefont
  {J.~C.}\ \bibnamefont {Hamley}}, \bibinfo {author} {\bibfnamefont
  {N.}~\bibnamefont {Denny}}, \bibinfo {author} {\bibfnamefont
  {G.}~\bibnamefont {Bloye}}, \bibinfo {author} {\bibfnamefont
  {W.}~\bibnamefont {Zhang}}, \bibinfo {author} {\bibfnamefont
  {L.}~\bibnamefont {Hentges}}, \bibinfo {author} {\bibfnamefont
  {R.}~\bibnamefont {Doll}}, \bibinfo {author} {\bibfnamefont {Y.}~\bibnamefont
  {Wei}}, \bibinfo {author} {\bibfnamefont {M.}~\bibnamefont {Maresca}},
  \bibinfo {author} {\bibfnamefont {E.}~\bibnamefont {Dimitrova}}, \bibinfo
  {author} {\bibfnamefont {L.}~\bibnamefont {Pytowski}}, \bibinfo {author}
  {\bibfnamefont {E.~A.}\ \bibnamefont {Tunnacliffe}}, \bibinfo {author}
  {\bibfnamefont {M.}~\bibnamefont {Kassouf}}, \bibinfo {author} {\bibfnamefont
  {D.}~\bibnamefont {Higgs}}, \bibinfo {author} {\bibfnamefont
  {E.}~\bibnamefont {de~Wit}}, \bibinfo {author} {\bibfnamefont {R.~J.}\
  \bibnamefont {Klose}}, \bibinfo {author} {\bibfnamefont {C.}~\bibnamefont
  {Schermelleh}}, \bibinfo {author} {\bibfnamefont {R.}~\bibnamefont
  {Collepardo-Guevara}}, \bibinfo {author} {\bibfnamefont {T.~A.}\ \bibnamefont
  {Milne}},\ and\ \bibinfo {author} {\bibfnamefont {J.~O.}\ \bibnamefont
  {Davies}},\ }\href {https://doi.org/10.1016/j.cell.2025.10.013} {\bibfield
  {journal} {\bibinfo  {journal} {Cell}\ }\textbf {\bibinfo {volume} {186}},\
  \bibinfo {pages} {4735} (\bibinfo {year} {2025})}\BibitemShut {NoStop}%
\end{thebibliography}%


\begin{thebibliography}{7}%
\makeatletter
\providecommand \@ifxundefined [1]{%
 \@ifx{#1\undefined}
}%
\providecommand \@ifnum [1]{%
 \ifnum #1\expandafter \@firstoftwo
 \else \expandafter \@secondoftwo
 \fi
}%
\providecommand \@ifx [1]{%
 \ifx #1\expandafter \@firstoftwo
 \else \expandafter \@secondoftwo
 \fi
}%
\providecommand \natexlab [1]{#1}%
\providecommand \enquote  [1]{``#1''}%
\providecommand \bibnamefont  [1]{#1}%
\providecommand \bibfnamefont [1]{#1}%
\providecommand \citenamefont [1]{#1}%
\providecommand \href@noop [0]{\@secondoftwo}%
\providecommand \href [0]{\begingroup \@sanitize@url \@href}%
\providecommand \@href[1]{\@@startlink{#1}\@@href}%
\providecommand \@@href[1]{\endgroup#1\@@endlink}%
\providecommand \@sanitize@url [0]{\catcode `\\12\catcode `\$12\catcode
  `\&12\catcode `\#12\catcode `\^12\catcode `\_12\catcode `\%12\relax}%
\providecommand \@@startlink[1]{}%
\providecommand \@@endlink[0]{}%
\providecommand \url  [0]{\begingroup\@sanitize@url \@url }%
\providecommand \@url [1]{\endgroup\@href {#1}{\urlprefix }}%
\providecommand \urlprefix  [0]{URL }%
\providecommand \Eprint [0]{\href }%
\providecommand \doibase [0]{https://doi.org/}%
\providecommand \selectlanguage [0]{\@gobble}%
\providecommand \bibinfo  [0]{\@secondoftwo}%
\providecommand \bibfield  [0]{\@secondoftwo}%
\providecommand \translation [1]{[#1]}%
\providecommand \BibitemOpen [0]{}%
\providecommand \bibitemStop [0]{}%
\providecommand \bibitemNoStop [0]{.\EOS\space}%
\providecommand \EOS [0]{\spacefactor3000\relax}%
\providecommand \BibitemShut  [1]{\csname bibitem#1\endcsname}%
\let\auto@bib@innerbib\@empty
\bibitem [{\citenamefont {Joseph}\ \emph {et~al.}(2021)\citenamefont {Joseph},
  \citenamefont {Reinhardt}, \citenamefont {Aguirre}, \citenamefont {Chew},
  \citenamefont {Russell}, \citenamefont {Espinosa}, \citenamefont {Garaizar},\
  and\ \citenamefont {Collepardo-Guevara}}]{Mpipi2021}%
  \BibitemOpen
  \bibfield  {author} {\bibinfo {author} {\bibfnamefont {J.~A.}\ \bibnamefont
  {Joseph}}, \bibinfo {author} {\bibfnamefont {A.}~\bibnamefont {Reinhardt}},
  \bibinfo {author} {\bibfnamefont {A.}~\bibnamefont {Aguirre}}, \bibinfo
  {author} {\bibfnamefont {P.~Y.}\ \bibnamefont {Chew}}, \bibinfo {author}
  {\bibfnamefont {K.~O.}\ \bibnamefont {Russell}}, \bibinfo {author}
  {\bibfnamefont {J.~R.}\ \bibnamefont {Espinosa}}, \bibinfo {author}
  {\bibfnamefont {A.}~\bibnamefont {Garaizar}},\ and\ \bibinfo {author}
  {\bibfnamefont {R.}~\bibnamefont {Collepardo-Guevara}},\ }\href@noop {}
  {\bibfield  {journal} {\bibinfo  {journal} {Nature Computational Science}\
  }\textbf {\bibinfo {volume} {1}},\ \bibinfo {pages} {732} (\bibinfo {year}
  {2021})}\BibitemShut {NoStop}%
\bibitem [{\citenamefont {Chambers}\ \emph {et~al.}(2003)\citenamefont
  {Chambers}, \citenamefont {Colby}, \citenamefont {Robertson}, \citenamefont
  {Nichols}, \citenamefont {Lee}, \citenamefont {Tweedie},\ and\ \citenamefont
  {Smith}}]{Chambers2003}%
  \BibitemOpen
  \bibfield  {author} {\bibinfo {author} {\bibfnamefont {I.}~\bibnamefont
  {Chambers}}, \bibinfo {author} {\bibfnamefont {D.}~\bibnamefont {Colby}},
  \bibinfo {author} {\bibfnamefont {M.}~\bibnamefont {Robertson}}, \bibinfo
  {author} {\bibfnamefont {J.}~\bibnamefont {Nichols}}, \bibinfo {author}
  {\bibfnamefont {S.}~\bibnamefont {Lee}}, \bibinfo {author} {\bibfnamefont
  {S.}~\bibnamefont {Tweedie}},\ and\ \bibinfo {author} {\bibfnamefont
  {A.}~\bibnamefont {Smith}},\ }\href@noop {} {\bibfield  {journal} {\bibinfo
  {journal} {Cell}\ }\textbf {\bibinfo {volume} {113}},\ \bibinfo {pages} {643}
  (\bibinfo {year} {2003})}\BibitemShut {NoStop}%
\bibitem [{\citenamefont {Thompson}\ \emph {et~al.}(2022)\citenamefont
  {Thompson}, \citenamefont {Aktulga}, \citenamefont {Berger}, \citenamefont
  {Bolintineanu}, \citenamefont {Brown}, \citenamefont {Crozier}, \citenamefont
  {in~'t Veld}, \citenamefont {Kohlmeyer}, \citenamefont {Moore}, \citenamefont
  {Nguyen}, \citenamefont {Shan}, \citenamefont {Stevens}, \citenamefont
  {Tranchida}, \citenamefont {Trott},\ and\ \citenamefont
  {Plimpton}}]{LAMMPS2022}%
  \BibitemOpen
  \bibfield  {author} {\bibinfo {author} {\bibfnamefont {A.~P.}\ \bibnamefont
  {Thompson}}, \bibinfo {author} {\bibfnamefont {H.~M.}\ \bibnamefont
  {Aktulga}}, \bibinfo {author} {\bibfnamefont {R.}~\bibnamefont {Berger}},
  \bibinfo {author} {\bibfnamefont {D.~S.}\ \bibnamefont {Bolintineanu}},
  \bibinfo {author} {\bibfnamefont {W.~M.}\ \bibnamefont {Brown}}, \bibinfo
  {author} {\bibfnamefont {P.~S.}\ \bibnamefont {Crozier}}, \bibinfo {author}
  {\bibfnamefont {P.~J.}\ \bibnamefont {in~'t Veld}}, \bibinfo {author}
  {\bibfnamefont {A.}~\bibnamefont {Kohlmeyer}}, \bibinfo {author}
  {\bibfnamefont {S.~G.}\ \bibnamefont {Moore}}, \bibinfo {author}
  {\bibfnamefont {T.~D.}\ \bibnamefont {Nguyen}}, \bibinfo {author}
  {\bibfnamefont {R.}~\bibnamefont {Shan}}, \bibinfo {author} {\bibfnamefont
  {M.~J.}\ \bibnamefont {Stevens}}, \bibinfo {author} {\bibfnamefont
  {J.}~\bibnamefont {Tranchida}}, \bibinfo {author} {\bibfnamefont
  {C.}~\bibnamefont {Trott}},\ and\ \bibinfo {author} {\bibfnamefont {S.~J.}\
  \bibnamefont {Plimpton}},\ }\href {https://doi.org/10.1016/j.cpc.2021.108171}
  {\bibfield  {journal} {\bibinfo  {journal} {Comp. Phys. Comm.}\ }\textbf
  {\bibinfo {volume} {271}},\ \bibinfo {pages} {108171} (\bibinfo {year}
  {2022})}\BibitemShut {NoStop}%
\bibitem [{\citenamefont {Mullin}\ \emph {et~al.}(2008)\citenamefont {Mullin},
  \citenamefont {Yates}, \citenamefont {Rowe}, \citenamefont {Nijmeijer},
  \citenamefont {Colby}, \citenamefont {Barlow}, \citenamefont {Walkinshaw},\
  and\ \citenamefont {Chambers}}]{Mullin2008}%
  \BibitemOpen
  \bibfield  {author} {\bibinfo {author} {\bibfnamefont {N.}~\bibnamefont
  {Mullin}}, \bibinfo {author} {\bibfnamefont {A.}~\bibnamefont {Yates}},
  \bibinfo {author} {\bibfnamefont {A.}~\bibnamefont {Rowe}}, \bibinfo {author}
  {\bibfnamefont {B.}~\bibnamefont {Nijmeijer}}, \bibinfo {author}
  {\bibfnamefont {D.}~\bibnamefont {Colby}}, \bibinfo {author} {\bibfnamefont
  {P.}~\bibnamefont {Barlow}}, \bibinfo {author} {\bibfnamefont
  {M.}~\bibnamefont {Walkinshaw}},\ and\ \bibinfo {author} {\bibfnamefont
  {I.}~\bibnamefont {Chambers}},\ }\href {https://doi.org/10.1042/BJ20080134}
  {\bibfield  {journal} {\bibinfo  {journal} {Biochemical Journal}\ }\textbf
  {\bibinfo {volume} {411}},\ \bibinfo {pages} {227} (\bibinfo {year}
  {2008})}\BibitemShut {NoStop}%
\bibitem [{\citenamefont {Burt}\ \emph {et~al.}(2024)\citenamefont {Burt},
  \citenamefont {Toader}, \citenamefont {Warshamanage}, \citenamefont {von
  Kügelgen}, \citenamefont {Pyle}, \citenamefont {Zivanov}, \citenamefont
  {Kimanius}, \citenamefont {Bharat},\ and\ \citenamefont {Scheres}}]{relion}%
  \BibitemOpen
  \bibfield  {author} {\bibinfo {author} {\bibfnamefont {A.}~\bibnamefont
  {Burt}}, \bibinfo {author} {\bibfnamefont {B.}~\bibnamefont {Toader}},
  \bibinfo {author} {\bibfnamefont {R.}~\bibnamefont {Warshamanage}}, \bibinfo
  {author} {\bibfnamefont {A.}~\bibnamefont {von Kügelgen}}, \bibinfo {author}
  {\bibfnamefont {E.}~\bibnamefont {Pyle}}, \bibinfo {author} {\bibfnamefont
  {J.}~\bibnamefont {Zivanov}}, \bibinfo {author} {\bibfnamefont
  {D.}~\bibnamefont {Kimanius}}, \bibinfo {author} {\bibfnamefont {T.~A.~M.}\
  \bibnamefont {Bharat}},\ and\ \bibinfo {author} {\bibfnamefont {S.~H.~W.}\
  \bibnamefont {Scheres}},\ }\href
  {https://doi.org/https://doi.org/10.1002/2211-5463.13873} {\bibfield
  {journal} {\bibinfo  {journal} {FEBS Open Bio}\ }\textbf {\bibinfo {volume}
  {14}},\ \bibinfo {pages} {1788} (\bibinfo {year} {2024})}\BibitemShut
  {NoStop}%
\bibitem [{\citenamefont {Mullin}\ \emph {et~al.}(2017)\citenamefont {Mullin},
  \citenamefont {Gagliardi}, \citenamefont {Khoa}, \citenamefont {Colby},
  \citenamefont {Hall-Ponsele}, \citenamefont {Rowe},\ and\ \citenamefont
  {Chambers}}]{Mullin2017}%
  \BibitemOpen
  \bibfield  {author} {\bibinfo {author} {\bibfnamefont {N.~P.}\ \bibnamefont
  {Mullin}}, \bibinfo {author} {\bibfnamefont {A.}~\bibnamefont {Gagliardi}},
  \bibinfo {author} {\bibfnamefont {L.~T.~P.}\ \bibnamefont {Khoa}}, \bibinfo
  {author} {\bibfnamefont {D.}~\bibnamefont {Colby}}, \bibinfo {author}
  {\bibfnamefont {E.}~\bibnamefont {Hall-Ponsele}}, \bibinfo {author}
  {\bibfnamefont {A.~J.}\ \bibnamefont {Rowe}},\ and\ \bibinfo {author}
  {\bibfnamefont {I.}~\bibnamefont {Chambers}},\ }\href
  {https://doi.org/https://doi.org/10.1016/j.jmb.2016.12.001} {\bibfield
  {journal} {\bibinfo  {journal} {Journal of Molecular Biology}\ }\textbf
  {\bibinfo {volume} {429}},\ \bibinfo {pages} {1544} (\bibinfo {year}
  {2017})}\BibitemShut {NoStop}%
\bibitem [{\citenamefont {Jauch}\ \emph {et~al.}(2008)\citenamefont {Jauch},
  \citenamefont {Ng}, \citenamefont {Saikatendu}, \citenamefont {Stevens},\
  and\ \citenamefont {Kolatkar}}]{Jauch2008}%
  \BibitemOpen
  \bibfield  {author} {\bibinfo {author} {\bibfnamefont {R.}~\bibnamefont
  {Jauch}}, \bibinfo {author} {\bibfnamefont {C.~K.~L.}\ \bibnamefont {Ng}},
  \bibinfo {author} {\bibfnamefont {K.~S.}\ \bibnamefont {Saikatendu}},
  \bibinfo {author} {\bibfnamefont {R.~C.}\ \bibnamefont {Stevens}},\ and\
  \bibinfo {author} {\bibfnamefont {P.~R.}\ \bibnamefont {Kolatkar}},\ }\href
  {https://doi.org/https://doi.org/10.1016/j.jmb.2007.11.091} {\bibfield
  {journal} {\bibinfo  {journal} {Journal of Molecular Biology}\ }\textbf
  {\bibinfo {volume} {376}},\ \bibinfo {pages} {758} (\bibinfo {year}
  {2008})}\BibitemShut {NoStop}%
\end{thebibliography}%
\end{document}


\title{Supplementary information: Modulation of DNA rheology by a transcription factor that forms aging microgels}

\affiliation{Centre for Regenerative Medicine, Institute for Regeneration and Repair, 5 Little France Drive, Edinburgh EH16 4UU, Scotland}
\affiliation{School of Physics and Astronomy, University of Edinburgh, Edinburgh Eh9 3FD, Scotland}

\author{Yair Augusto Guti\'{e}rrez Fosado}
\thanks{Joint first author}
\affiliation{School of Physics and Astronomy, University of Edinburgh, Edinburgh Eh9 3FD, Scotland}

\author{Abbie Guild}
\affiliation{Centre for Regenerative Medicine, Institute for Regeneration and Repair, 5 Little France Drive, Edinburgh EH16 4UU, Scotland}
\affiliation{Institute for Stem Cell Research, School of Biological Sciences, University of Edinburgh, 5 Little France Drive, Edinburgh EH16 4UU, Scotland}

\author{Nicholas Mullin}
\affiliation{Centre for Regenerative Medicine, Institute for Regeneration and Repair, 5 Little France Drive, Edinburgh EH16 4UU, Scotland}
\affiliation{Institute for Stem Cell Research, School of Biological Sciences, University of Edinburgh, 5 Little France Drive, Edinburgh EH16 4UU, Scotland}

\author{Laura Spagnolo}
\affiliation{School of Molecular Biosciences, University of Glasgow, University Avenue, Glasgow G12 8QQ, Scotland}

\author{Ian Chambers}
\thanks{corresponding author, ichambers@ed.ac.uk}
\affiliation{Centre for Regenerative Medicine, Institute for Regeneration and Repair, 5 Little France Drive, Edinburgh EH16 4UU, Scotland}
\affiliation{Institute for Stem Cell Research, School of Biological Sciences, University of Edinburgh, 5 Little France Drive, Edinburgh EH16 4UU, Scotland}

\author{Davide Michieletto}
\thanks{corresponding author, davide.michieletto@ed.ac.uk}
\affiliation{School of Physics and Astronomy, University of Edinburgh, Edinburgh Eh9 3FD, Scotland}
\affiliation{MRC Human Genetics Unit, Institute of Genetics and Cancer, University of Edinburgh, Scotland}
\affiliation{International Institute for Sustainability with Knotted Chiral Meta Matter (WPI-SKCM$^2$), Hiroshima University, Higashi-Hiroshima, Hiroshima 739-8526, Japan}

\maketitle

\section{The Mpipi model}
To simulate NANOG molecules, we employed the Mpipi coarse-grained model~\cite{Mpipi2021}. In Mpipi, each amino acid is represented as a single isotropic bead characterized by its mass, diameter and charge. The total potential energy is expressed as

\begin{equation}
V_{\mathrm{Mpipi}} = \sum_{\mathrm{bonds}} V_{\mathrm{bond}} + \sum_{\mathrm{pairs}} (V_{\mathrm{elec}} + V_{\mathrm{pair}}),
\label{eq:Umpipi}
\end{equation}

\noindent where $V_{\mathrm{bond}}$ is a harmonic term enforcing chain connectivity, $V_{\mathrm{elec}}$ is a screened Coulomb potential described by the Debye--Hückel formalism, and $V_{\mathrm{pair}}$ is a Wang--Frenkel potential that governs non-bonded residue interactions.

Model parameters were derived from a combination of all-atom potential of mean force (PMF) calculations at 150~mM salt concentration and sequence statistics from bioinformatics datasets. This hybrid parameterization enables Mpipi to capture key residue-specific noncovalent interactions, including hydrophobic, cation--($\pi$), and ($\pi$)--($\pi$) interactions. Notably, the model distinguishes between arginine and lysine residues, reflecting their distinct propensities for electrostatic and aromatic interactions.

Mpipi has been validated against a range of experimental observables, including protein radii of gyration and phase diagrams, and has demonstrated robust predictive accuracy for both protein--RNA and protein--protein phase separation.

\section{Details of NANOG coarse-grained simulations}
The mouse NANOG protein comprises $N=305$ amino acids (sequence shown in Table~\ref{table:seq}) and contains three major domains: the N-terminal domain (ND), the homeodomain (HD), and the C-terminal domain (CD). The CD itself is subdivided into CD1, the tryptophan repeat (WR), and CD2~\cite{Chambers2003} (see Fig.~\ref{M-fig:aging_protein}(a) in the main text). We note that, based on sequence alone, NANOG is expected to be weakly negatively charged. Additionally, the WR region contains no charged amino acids. Thus, the original parameterization~\cite{Mpipi2021} of the Mpipi model is expected to provide a reliable description of the system.

\subsection{Computation of radius of gyration in dilute conditions}
Coarse-grained molecular dynamics simulations were performed for a single polymer of either wild-type NANOG or the W10A mutant under dilute conditions. Each molecule was placed in a cubic simulation box of side length $L = 60$ nm and evolved in the NVT ensemble using the LAMMPS package~\cite{LAMMPS2022} with implicit solvent (Langevin dynamics) at a temperature of $T = 300$ K. The equations of motion were integrated with a time step of $10$ fs for a total simulation time of $5~\mu$s ($5 \times 10^8$ steps). The radius of gyration ($R_g$) was calculated as

\begin{equation}
R_g^2 = \frac{1}{M_t} \sum_{i=1}^{N} m_i \lvert \mathbf{r}_i - \mathbf{r}_{\mathrm{com}} \rvert^2 ,    
\end{equation}

\noindent where $M_t$ is the total mass of the polymer, $\mathbf{r}_{\mathrm{com}}$ is the polymer’s center of mass, and $m_i$, $\mathbf{r}_i$ are the mass and position of the $i$-th bead, respectively. $R_g$ values were recorded every $10^5$ time steps (1~ns). The first $1~\mu\mathrm{s}$ of each trajectory was excluded from the analysis to ensure equilibration. The average radius of gyration obtained was $R_g = 3.9 \pm 0.9$ nm for wild-type NANOG and $R_g = 4.8 \pm 1.2$ nm for the W10A mutant.

\begin{table*}[t]
\centering  
\resizebox{1.0\textwidth}{!}{%
\begin{tabular}{|c|c|c|}
\hline
\textbf{Region} & \textbf{Length} & \textbf{Residue sequence} \\
\hline
ND  & 95 & MSVGLPGPHSLPSSEEASNSGNASSMPAVFHPENYSCLQGSATEMLCTEAASPRPSSEDLPLQGSPDSSTSPKQKLSSPEADKGPEEEENKVLAR \\
HD  & 60 & KQKMRTVFSQAQLCALKDRFQKQKYLSLQQMQELSSILNLSYKQVKTWFQNQRMKCKRWQ \\
CD1 & 42 & KNQWLKTSNGLIQKGSAPVEYPSIHCSYPQGYLVNASGSLSM \\
WR  & 50 & WGSQTWTNPTWSSQTWTNPTWNNQTWTNPTWSSQAWTAQSWNGQPWNAAP \\
CD2 & 58 & LHNFGEDFLQPYVQLQQNFSASDLEVNLEATRESHAHFSTPQALELFLNYSVTPPGEI \\
\hline
\end{tabular}
}
\caption{\textbf{NANOG sequence across distinct regions.} Amino acid sequences corresponding to the N-terminal domain (ND), homeodomain (HD), C-terminal domain 1 (CD1), tryptophan repeat (WR), and C-terminal domain 2 (CD2) are shown. The second column reports the number of amino acids in each region.}
\label{table:seq}
\end{table*}

\subsection{Cluster simulations of proteins}
Simulations of proteins were carried out using the Mpipi coarse-grained model. The simulation protocol was adapted from the direct-coexistence approach, a standard method for probing phase separation in biomolecular systems. In this setup, protein molecules are placed within an elongated simulation box that allows for the potential coexistence of dense and dilute phases, if present.

The simulation box was defined as a rectangular prism with its longest axis aligned along the $z$-direction. Periodic boundary conditions were applied in all three dimensions. To minimize finite-size effects and prevent artificial interactions between periodic images, the shorter box dimensions ($L_x$ and $L_y$) were set to at least two to three times the radius of gyration ($R_g$) of a single protein under dilute conditions. The box length $L_z$ was then adjusted to achieve an overall biomolecular concentration corresponding to an average density of approximately $0.1~\mathrm{g/cm^{3}}$.

We simulated $M = 50$ protein molecules in total. For wild-type NANOG, the simulation box dimensions were $L_x = L_y = 15$ nm and $L_z = 70$ nm, whereas for the W10A mutant the box size was $L_x = L_y = 18$ nm and $L_z = 53$ nm. Each system was first equilibrated at a high temperature ($T = 473$ K) for $5 \times 10^{6}$ timesteps (approximately $0.5~\mu$s) to prevent any attractive interactions between polymers and generate a uniform density distribution in the box. Following equilibration, the temperature was quenched to $T = 300$ K, and production simulations were performed in the NVT ensemble using a Langevin thermostat with a relaxation time of 5~ps and an integration timestep of 10~fs, for a total duration of $\sim 2 \times 10^{9}$ timesteps ($2~\mu$s) for NANOG and $1 \times 10^{9}$ timesteps for W10A. For each case we ran three independent replicas.

\begin{figure*}[t]
	\begin{center}
		\includegraphics[width=1.0\textwidth]{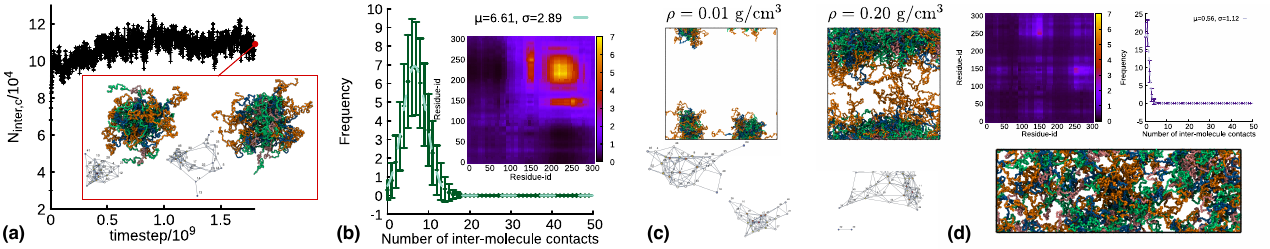}
		\caption{\textbf{NANOG forms self-limited clusters}. \textbf{(a)} Time evolution of the total number of intermolecule contacts. Inset shows a snapshot from equilibrated simulations of NANOG ($M = 50$, $N = 305$, $T = 300$ K, $\rho = 0.1~\mathrm{g/cm^3}$). Network diagrams show the molecules in each of the two clusters (29 and 21 molecules, respectively). \textbf{(b)} Main: distribution of the number of intermolecular contacts. Inset: residue-level intermolecular contact map. \textbf{(c)} Equilibrium configurations from simulations and the corresponding network diagrams. Left: $\rho = 0.01~\mathrm{g/cm^3}$, showing two clusters containing 29 and 21 molecules, respectively. Right: $\rho = 0.2~\mathrm{g/cm^3}$. \textbf{(d)} Data from simulations of the W10A mutant ($M = 50$, $N = 305$, $T = 300$ K, $\rho = 0.1~\mathrm{g/cm^3}$), comprising a representative snapshot, the distribution of intermolecular contacts, and the residue-level contact map.}
        \label{fig:simulations}
	\end{center}
    \vspace{-0.6cm}
\end{figure*}
\subsection{Data analysis}

To assess whether the system had reached equilibrium, we first monitored the total number of intermolecular contacts ($N_{\mathrm{inter,c}}$). Contacts were defined between residues from different proteins when their distance satisfied $r_{ij} \leq 14.5~\text{\AA}$, corresponding to the shortest cutoff distance of the Wang-Frenkel potential for non-charged particles in the Mpipi model~\cite{Mpipi2021}. The time evolution of $N_{\mathrm{inter,c}}$ is shown in Fig.~\ref{fig:simulations}(a). Based on this analysis, the initial $16~\mu$s of the NANOG simulation and the first $7.5~\mu$s of the W10A simulation were excluded from subsequent analysis.

After equilibration, the simulations revealed that NANOG molecules spontaneously assemble into self-limited clusters, each comprising a well-defined maximum size of approximately 30 proteins. A representative snapshot of the equilibrated system (inset of Fig.~\ref{fig:simulations}(a)) shows the formation of two stable clusters (see also Fig.~\ref{M-fig2}(a) of the main text). 

To characterize the internal organization of these assemblies, we computed the intermolecular contact map (Fig.~\ref{fig:simulations}(b), inset). Regions of high contact probability, shown in light colors, are concentrated within the WR segment (residues 195–268). A distinct band of interactions is also evident between the WR and the HD (residues 142–159), indicating frequent cross-domain associations. From the WR–WR interactions, we further obtained the distribution of intermolecular connections, defined as the number of NANOG molecules with which a given molecule forms WR-mediated contacts. This distribution follows a Gaussian profile, with an average of seven intermolecular contacts per NANOG molecule within individual clusters (Fig.~\ref{fig:simulations}(b)).

At a given timestep, the connectivity between molecules can be represented as a network diagram (see inset of Fig.~\ref{fig:simulations}(a)), where each vertex corresponds to a NANOG molecule and edges connect pairs of molecules that form WR-mediated contacts. A connected component is defined as a set of vertices in which any two molecules are linked by one or more connecting paths. Analysis of this network revealed that the system consistently organizes into two main connected components, corresponding to the two clusters observed in the simulations. Occasionally, a few molecules transiently detach from these clusters, forming isolated monomers or dimers. This behaviour was quantified by monitoring the time evolution of the number of molecules in each connected component (see Fig.~\ref{M-fig2}(c) in the main text). Notably, across all independent replicas, the size of the largest connected component never exceeded 30 molecules, confirming that NANOG forms self-limited clusters.

To examine this self-limiting behaviour and the stability of clusters, we performed a controlled test in which the two NANOG clusters were driven to collide by applying external forces. After the clusters merged, the forces were removed and the system was allowed to relax for $20~\mu\mathrm{s}$. Remarkably, the system spontaneously reorganized into two clusters again, with proteins from the initial aggregates redistributed between the new ones (Fig.~\ref{M-fig2}(e) of the main text).

In classical phase separation, systems prepared within the coexistence region spontaneously separate into coexisting high- and low-density phases. Although the densities of these phases remain constant, the number of molecules in each phase, and consequently the size of the condensed domains, typically varies with the overall concentration. This naturally raises the question of whether the cluster size observed here depends on box geometry or initial concentration. As shown in Fig.~\ref{fig:simulations}(c, left), this is not the case: a system initialized at tenfold lower density ($\rho = 0.01~\mathrm{g/cm^3}$) in a cubic box still formed two clusters, supporting the self-limited nature of the assembly. At higher densities ($\rho = 0.2~\mathrm{g/cm^3}$; Fig.~\ref{fig:simulations}(c, right)), however, the system transitioned from discrete clusters to a percolating network in which a single aggregate spanned the entire box, marking the onset of gel-like behaviour, consistent with the sample shown in Fig.~\ref{M-fig:aging_protein}(c) of the main text.

Finally, we performed simulations of the W10A mutant. Substituting the tryptophan residues with alanine in the WR region resulted in a uniform distribution of molecules throughout the simulation box (Fig.~\ref{fig:simulations}(d)). No evidence of cluster formation was observed, underscoring the essential role of the tryptophan residues in mediating NANOG self-association, consistent with previous experimental observations~\cite{Mullin2008}.

\section{CryoEM}
3 $\mu$l of purified recombinant NANOG (see next section for purification method) at a concentration of 1 mg/ml were spotted on Quantifoil 2/2 grids with blotting for 3.5 seconds and blotted on a Vitrobot instrument with a blot force of 4. 6407 multi-frame movies in total were collected on a JEOL CRYOARM300 microscope equipped with a DE64 detector. The data collection was performed using Serial EM in linear mode using a pixel size of 1 \AA/pixel and 58 e-/\AA$^2$. The movies were motion corrected using MotionCorr and the CTF estimation was carried out using the CTFFIND 4.1 function in RELION 5.0~\cite{relion}. 1,726 movies were selected based on motion correction and CTF parameters. 1,036 particles were picked from the micrographs manually, extracted using a box size of 384x384 pixel and coarsened to 64x64 pixels prior to 2D classification. These class averages were then used as references for referenced autopicking of the data set producing a particle set of 1,048,156 particles. Following extraction, classification and particle selection in Relion 5.0, a final dataset of 121,110 particles was classified into 50 classes using a 240 \AA~ mask.


\section{Purification of recombinant NANOG}
Full-length Nanog, W10A and N51A were expressed in pET15b and purified according to established procedures described in Refs.~\cite{Mullin2008} and ~\cite{Mullin2017}. We briefly summarize the methods here.

\subsection{Purification of recombinant NANOG}
BL21(DE3) bacteria were transformed with NANOG-carrying gene under the control of an inducible promoter. The over-expression was triggered by 1mM isopropyl $\beta$-d-1-thiogalactopyranoside (IPTG) and grown for 3 hours at 37$^\circ$C. See Ref.~\cite{Mullin2017} for more details.

The protein was then purified in the presence of urea using nickel resin columns. Purified protein was concentrated in Vivaspin devices as in Ref.~\cite{Mullin2017}.
The protein was kept in denaturing buffer (25mM Hepes pH7.6, 150mM NaCl, 7M urea,
10mM imidazole, 10mM beta-mercaptoethanol). The protein was then run on an SDS-PAGE gel to confirm its purity and its concentration was measured by Bradford assay. 

\begin{figure*}[t!]
	\begin{center}
    \includegraphics[width=0.6\textwidth]{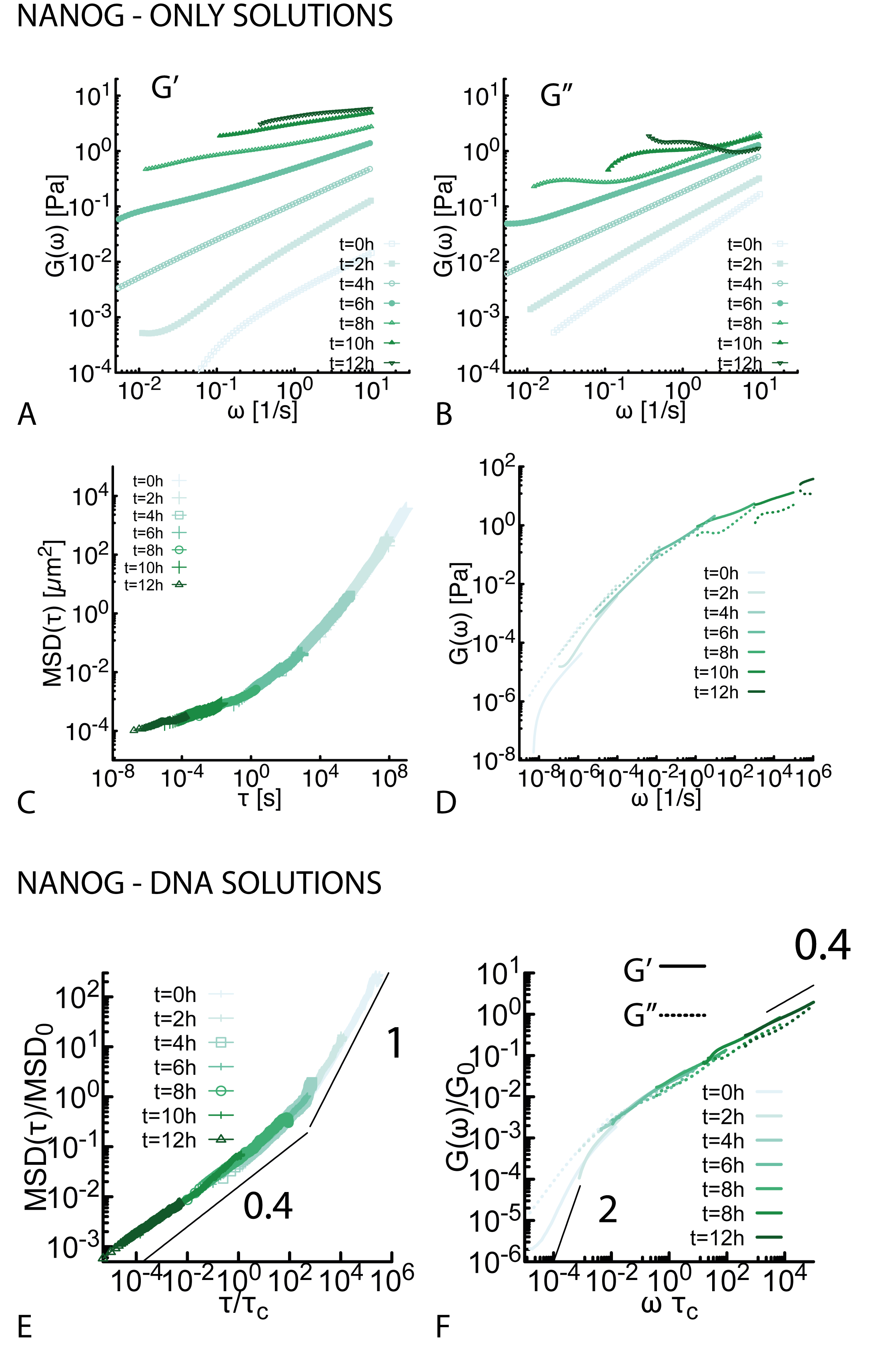}
		\caption{\textbf{Rheology analysis of NANOG solutions and NANOG-DNA mixtures}. \textbf{A-B} Time-resolved elastic and viscous moduli during the aging process. \textbf{C-D} Time-cure superpositions of MSDs and shear moduli of NANOG fluids during aging. \textbf{E-F} Time-cure superposition of MSDs and shear moduli of NANOG-DNA mixtures during aging. }
        \label{fig:aging_superpos}
	\end{center}
    \vspace{-0.6cm}
\end{figure*}

Concentrated protein was refolded by diluting denatured protein (10 mg/ml) in refolding buffer (10mM Hepes pH7.6, 50mM KCl, 10mM NaCl, 0.4mM EDTA, 10$\%$ glycerol) 1 in 10 (final 1 mg/ml) and incubating at room temperature for 30 minutes. Protein was stored refolded at 4$^\circ$C for up to 2 weeks and diluted in refolding buffer as needed.

\subsection{Electromobility shift assay (EMSA)}
The DNA probe used was a IRDye 680nm labelled 26bp sequence from the Tcf3 promoter~\cite{Jauch2008} (\textsf{TAAAC\hspace{0pt}CTGTT\hspace{0pt}AATGG\hspace{0pt}GAGCG\hspace{0pt}CATTG}). Final protein concentrations were as indicated in the figure. Samples were incubated for 30 min at room temp and analyzed on 1.5\% agarose TBE gels. The gel was imaged at 700nm.


\subsection{Additional Rheology Plots}
In this section we report additional rheology analysis. In Fig.~\ref{fig:aging_superpos} we report the time-cure superposition curves for the MSDs and $G^\prime$ and $G^{\prime \prime}$ for solutions of WT NANOG and DNA-NANOG mixtures. In the time-cure superposition plots for the MSDs, each curve at different ``curing time'' (aging time $t_a$) is shifted by a lagtime $\tau_c(t_a)$ and MSD value $m_c(t_a)$. The corresponding shear moduli are shifted in frequency by $\omega_c (t_a) = 1/\tau_c (t_a)$ and by a value $G_c(t_a)$.

\bibliography{bibliography}